\pdfoutput=1
\documentclass[apj]{emulateapj}
\usepackage{amssymb,graphicx}
\usepackage{epsfig}
\usepackage[usenames]{color}

\newcommand{\beqn}{\begin{eqnarray}}
\newcommand{\eeqn}{\end{eqnarray}}
\newcommand{\beq}{\begin{equation}}
\newcommand{\eeq}{\end{equation}}

\begin{document}

\title{Gravitational waves from the collision of tidally disrupted stars with massive black holes}
\author{William E.\ East}
\email{weast@stanford.edu}
\affil{
Kavli Institute for Particle Astrophysics and Cosmology, Stanford University, SLAC National Accelerator Laboratory, Menlo Park, CA 94025, USA
}

\begin{abstract}
We use simulations of hydrodynamics coupled with full general relativity to investigate the gravitational waves produced by a star colliding with a massive black hole 
when the star's tidal disruption radius lies far outside of the black hole horizon.
We consider both main-sequence and white-dwarf compaction stars, and nonspinning black holes, as well as those with near-extremal spin.
We study the regime in between where the star can be accurately modeled by a point particle, and where tidal effects completely 
suppress the gravitational wave signal.
We find that nonnegligible gravitational waves can be produced even when the star is strongly affected by tidal forces, as well
as when it collides with large angular momentum. 
We discuss the implications that these results have for the potential observation of gravitational waves from these sources with future detectors.
\end{abstract}


\maketitle

\section{Introduction}
\label{intro}

One exciting prospect for upcoming transient surveys is the possibility of detecting numerous tidal disruption events
resulting from a star being pulled apart by a massive black hole (BH) and then accreted.  There have already been 
a number of candidate observations of such events across the electromagnetic spectrum~\citep{1996A&A...309L..35B,1999A&A...349L..45K,2003ApJ...592...42G,2006ApJ...653L..25G,2008ApJ...676..944G,2009A&A...495L...9C,2011ApJ...741...73V,2011Sci...333..203B,2011Sci...333..199L,2011Natur.476..425Z,2012MNRAS.420.2684C,2012ApJ...753...77C,2012Natur.485..217G}, and with the large number of expected 
future observations~\citep{2011ApJ...741...73V}, there is hope that we will be able use such events to probe the strong-field gravity of massive BHs.
These transients could occur not only for main-sequence stars falling into supermassive BHs, but also possibly for white dwarfs
falling into intermediate-mass BHs~\citep{2010ApJ...712L...1I,2011ApJ...743..134K,2013ApJ...769...85S}.
However, complementing the events that produce these electromagnetic transients, there will be a population of stars launched on orbits without enough angular momentum to avoid
being swallowed by the BH.  To get a rough idea of the relative rates, we can assume that the stars filling the BH loss cone have negligible orbital energy
compared to their kinetic energy when orbiting near the BH, and that the differential rate $d\Gamma$ of stars with specific orbital angular momentum
$\tilde{L}$ obeys $d \Gamma \propto d(\tilde{L}^2)$~\citep{1976MNRAS.176..633F,2012PhRvD..85b4037K}, which holds in the so-called ``pinhole" regime.  
For a nonspinning BH, parabolic orbits with $\tilde{L}<\tilde{L}_{\rm cap}=4 M_{\rm BH}$ (where $M_{\rm BH}$ is the
mass of the BH and we use geometric units $G=c=1$ throughout, unless otherwise stated) fall into the BH horizon.  If we take the Newtonian estimate for the tidal disruption 
radius $r_{T} = R_*(M_{\rm BH}/M_{*})^{1/3}$ for a star of mass $M_{*}$ and radius $R_{*}$, then the angular momentum of a Newtonian parabolic orbit with a periapse equal to
$r_T$ is $\tilde{L}_{T}^2=2M^{4/3}_{\rm BH}R_*/M_*^{1/3}$.  Hence, the proportion of direct captures to those where the star is disrupted, but not captured, is given by
$f=(\tilde{L}_{T}^2/\tilde{L}_{\rm cap}^2-1)^{-1}$, where $\tilde{L}_{\rm cap}^2/\tilde{L}_{T}^2=8(M_{\rm BH}/M_*)^{2/3} (M_*/R_*)$.  Thus, for a supermassive BH with $M_{\rm BH} = 10^6$ $M_{\odot}$
and a star with solar mass and compaction, $M_*/R_*=2\times10^{-6}$, the ratio of collisions to disruption events is $\sim 0.2$.

A star colliding with a massive BH will most likely not produce a bright, fallback accretion-powered electromagnetic transient. 
However, such an event can produce gravitational waves (GWs).
Gravitational waves can also be produced by a star as it passes through periapse and is disrupted in a nonmerging encounter~\citep{2004ApJ...615..855K,2012ApJ...749..117H,2013PhRvD..87j4010C},
or potentially by the rebound of a tidally compressed star~\citep{2009ApJ...705..844G,Stone:2012uk}.
Gravitational waves produced by white dwarfs undergoing strong tidal interactions and mass transfer while orbiting massive BHs have also been considered~\citep{2010MNRAS.409L..25Z,2014ApJ...794....9M}.
However, such signals will, in general, be weaker than those resulting from collisions. 
In addition, the merger-ringdown signal from a collision event, if detected, directly reveals the mass and spin of the BH~\citep{Berti:2005ys}.
The ringdown frequency of supermassive BHs with masses $M_{\rm BH} \sim 10^5$--$10^8$ $M_{\odot}$ falls in the frequency band of a space-based \emph{LISA}-like GW instrument, while 
BHs with masses $M_{\rm BH}\lesssim 10^3$ $M_{\odot}$ fall in the band of LIGO~\citep{Abramovici:1992ah}, KAGRA~\citep{2012CQGra..29l4007S}, and other similar ground-based detectors.

However, there are several reasons such GW signals may be difficult to detect at large distances. 
To begin with, for an extreme-mass-ratio system the amplitude of the GW is proportional to the mass ratio
(for fixed total mass).  For extreme-mass-ratio inspirals of compact objects, considered a promising source for a \emph{LISA}-like GW detector~\citep{2007CQGra..24R.113A},
the system merges slowly through gravitational radiation reaction, and hence spends many wave cycles orbiting the BH. However, in the collision case, 
the timescale of the merger is just set by the mass of the BH and is much shorter.  Perhaps more importantly for the viability of these signals, 
there is the issue that as a star is pulled apart by
tidal forces, and its radius becomes comparable to that of the BH, the gravitational radiation it produces will be suppressed
compared to a point particle of the same mass~\citep{1982ApJ...257..283H}.    

It is the purpose of this paper to investigate the details of the GW signal produced by a star undergoing tidal disruption while
colliding with a massive BH.  
For stars whose radii are small compared to that of the BH, and whose tidal disruption radius does not lie outside the BH horizon, we expect
the gravitational waveform to be essentially identical to that of an equivalent point particle.  However, 
when tidal forces disperse the star's mass enough so that the different mass elements of the star no longer excite gravitational radiation 
from the BH coherently, the signal becomes significantly suppressed.  
In~\cite{1982ApJ...257..283H}, this was demonstrated by estimating the suppression of gravitational-wave energy due
to incoherence for the head-on collision of a star with a BH 
using BH perturbation theory. The star was modeled with a spherical dust cloud undergoing free fall
from the nominal tidal disruption radius of the star.  In the limit that the tidal radius goes to infinity, no gravitational
waves are produced.      

In this work, we are interested in studying the details of GW production
in the intermediate regime, between when the star can be treated as a point particle and tidal forces can be ignored 
and when GW production is minimal because of strong tidal forces (such as for the spherical dust cloud free-falling from infinity). 
We make use of recently developed techniques in the simulation of full general relativity coupled with hydrodynamics in the extreme-mass-ratio regime 
in order to take into account both the star's self-gravity and gravitational
perturbation, as well as the strong-field effects of the BH on the hydrodynamics of the star.
These methods allow us to self-consistently compute the full GW signal, including the ringdown of the perturbed BH.
Besides considering head-on collisions with nonspinning and spinning BHs, we also consider collisions with nonzero angular momentum.  Angular momentum
can significantly enhance the amplitude of the GW produced.  In fact, for a point particle, the energy in GWs formally diverges as
$\tilde{L}\to \tilde{L}_{\rm cap}$ since the corresponding geodesic circles the BH an infinite number of times at the innermost
stable orbit (thus signaling that back-reaction effects must be taken into account).  
For the star--BH system, it is thus of interest to investigate when finite size effects will be important in order to determine whether these
events might eventually constitute viable GW sources.   

The remainder of this paper is as follows.  We outline our numerical methods for accurately evolving the Einstein-hydrodynamic 
equations in Section~\ref{methods}, and describe the different cases we simulate in Section~\ref{cases}. We present the results of
these simulations in Section~\ref{results} where we find that nonnegligible GW radiation can be produced well into the regime where tidal effects are important,
both for head-on collisions, as well as those with significant angular momentum.  In Section~\ref{geo_model} we compare the simulations to a simple model
based on geodesics which we find explains the main features of the simulations well and can be used to generalize the results. We briefly comment on the detectability
of these GW signals in Section~\ref{detect}, and discuss the implications of these results and conclude in Section~\ref{conclusion}.

\section{Numerical methods and setup}
\label{methods}
In order to study the GW signal produced by the collision of a star with a massive BH, we 
perform simulations of hydrodynamics coupled with the Einstein-field equations using the code described in~\cite{code_paper}.
To accurately evolve the star's small contribution to the spacetime metric, without it being washed out
by truncation error from the dominant BH solution, we use the background error subtraction technique (BEST) 
developed in~\cite{best}.  BEST exploits the fact that the isolated BH solution is a known, exact solution to the
field equations, in order to explicitly subtract out its contribution to the truncation error for our particular numerical scheme.
This method was used to evolve star--BH systems with mass ratios up to $8\times10^6$ in~\cite{best}.  To minimize the advection error, we also perform the simulation in the rest frame of the star as
in~\cite{best}.  For the subset of cases considered below that are head-on collisions, we use the axisymmetry of the setup to restrict
the computational domain to the half-plane using a modified version of the Cartoon method~\citep{Alcubierre:1999ab} as described in~\cite{Pretorius:2004jg}.   

We model the stars as Tolman--Oppenheimer--Volkoff solutions with a gamma law equation state with $\Gamma=5/3$ (discussed in Section~\ref{cases}).  The initial
star profiles are taken to be polytropic, though we allow for shock heating during the evolution.  
We begin the simulation with the star outside its tidal disruption radius (at $d=200M_{\rm BH}$ and $100M_{\rm BH}$ for the head-on main-sequence star and white dwarf cases, respectively, and $d=1.2r_{T}$ for the three-dimensional cases) and choose the initial velocity of the star based on the geodesic of the BH spacetime with 
the corresponding orbital energy and angular momentum.  For all of the cases considered below, we assume that the star has zero
orbital energy, i.e., it is marginally unbound.

For the BH solution, we use harmonic coordinates~\citep{Cook:1997qc} for the nonspinning cases below; however, for the spinning
cases, we use Kerr--Schild coordinates since this solution is better behaved in the near-extremal-spin limit.  Consistent initial
data for the combined BH--star system is found by solving the constraint equations using the code described in~\cite{idsolve_paper}.

In order to measure the GW signal, we use the metric to calculate the Newman--Penrose scalar $\Psi_4$ at a radius
of $r=100M_{\rm BH}$ from the center of mass.  We decompose $\Psi_4$ into spin weight $-2$ spherical harmonics labeled by $\ell$ and $m$.  We note that for axisymmetric
cases only the $m=0$ components are nonzero.

\begin{figure}
\begin{center}
\includegraphics[width=3.6in,draft=false]{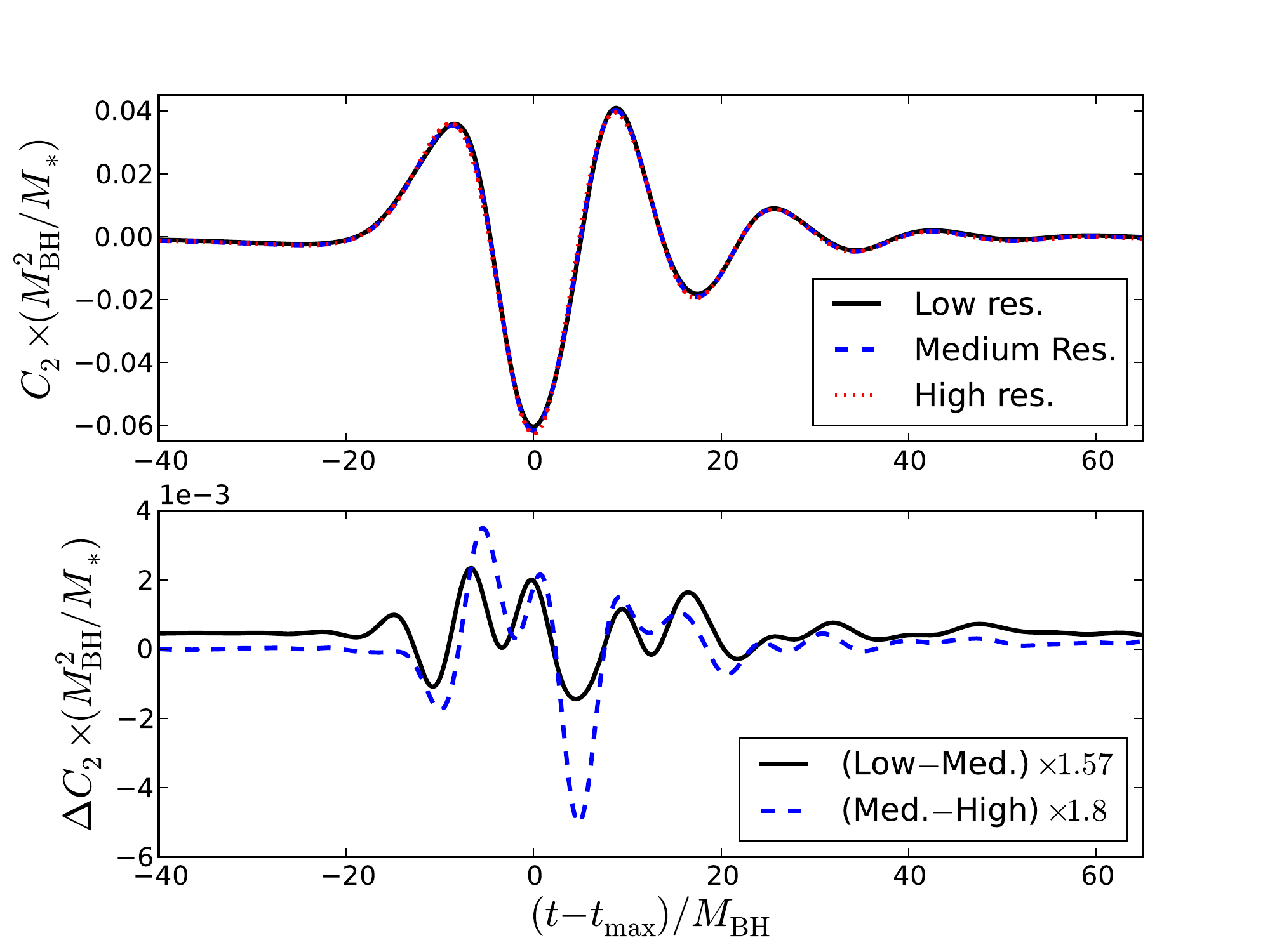}
\includegraphics[width=3.6in,draft=false]{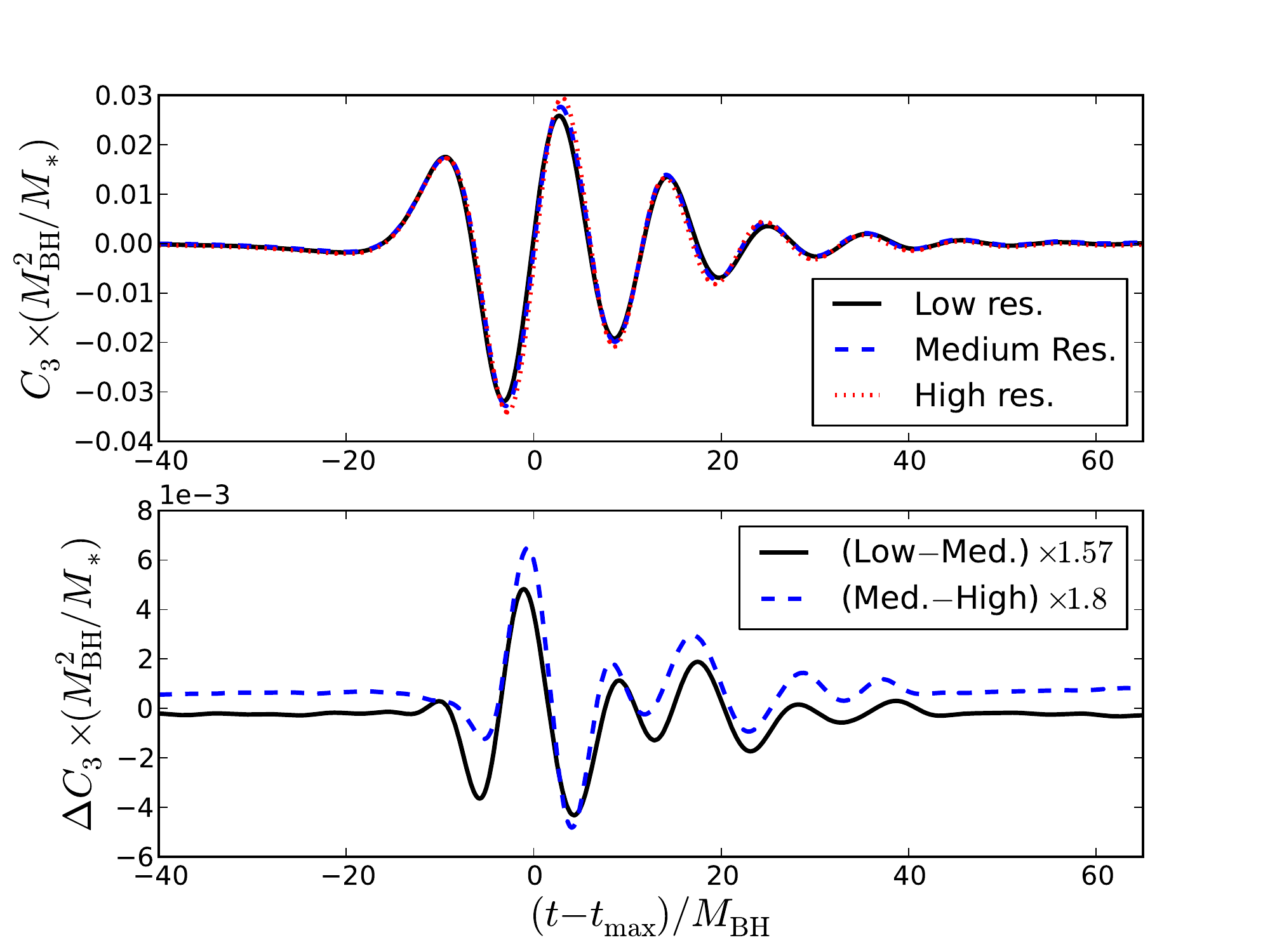}
\includegraphics[width=3.6in,draft=false]{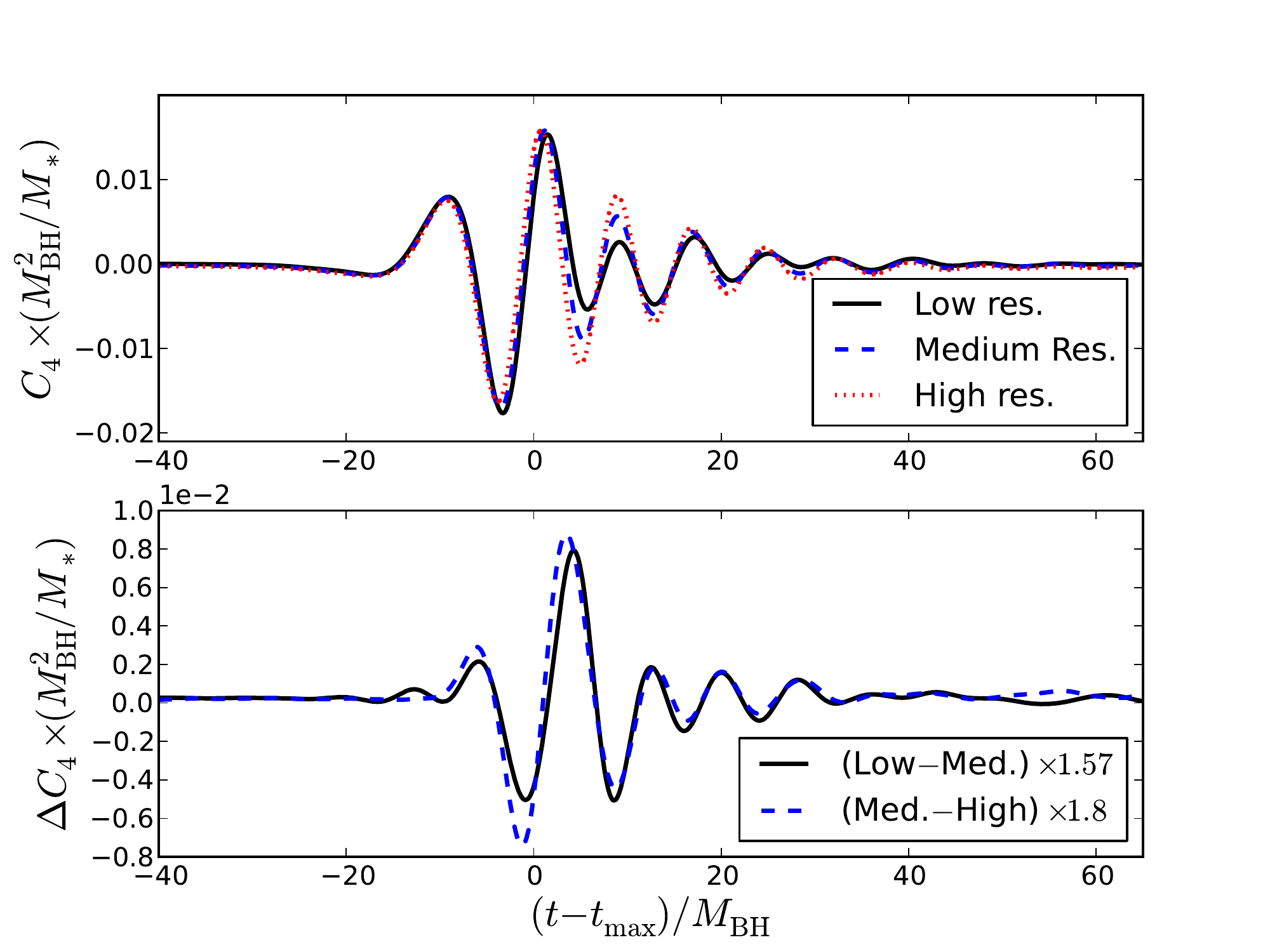}
\end{center}
\caption{
From top to bottom, the $\ell=2$, 3, and $4$ (with $m=0$) spin weight $-2$ spherical harmonics of $r\Psi_4$ from a 
nonspinning,  $M_{\rm BH}/M_*=10^6$, head-on collision simulation performed at three resolutions.
In each case, the lower panel shows the difference between the different resolutions, scaled assuming second-order convergence.
\label{conv_psi4_fig}
}
\end{figure}

\begin{figure}
\begin{center}
\includegraphics[width=3.6in,draft=false]{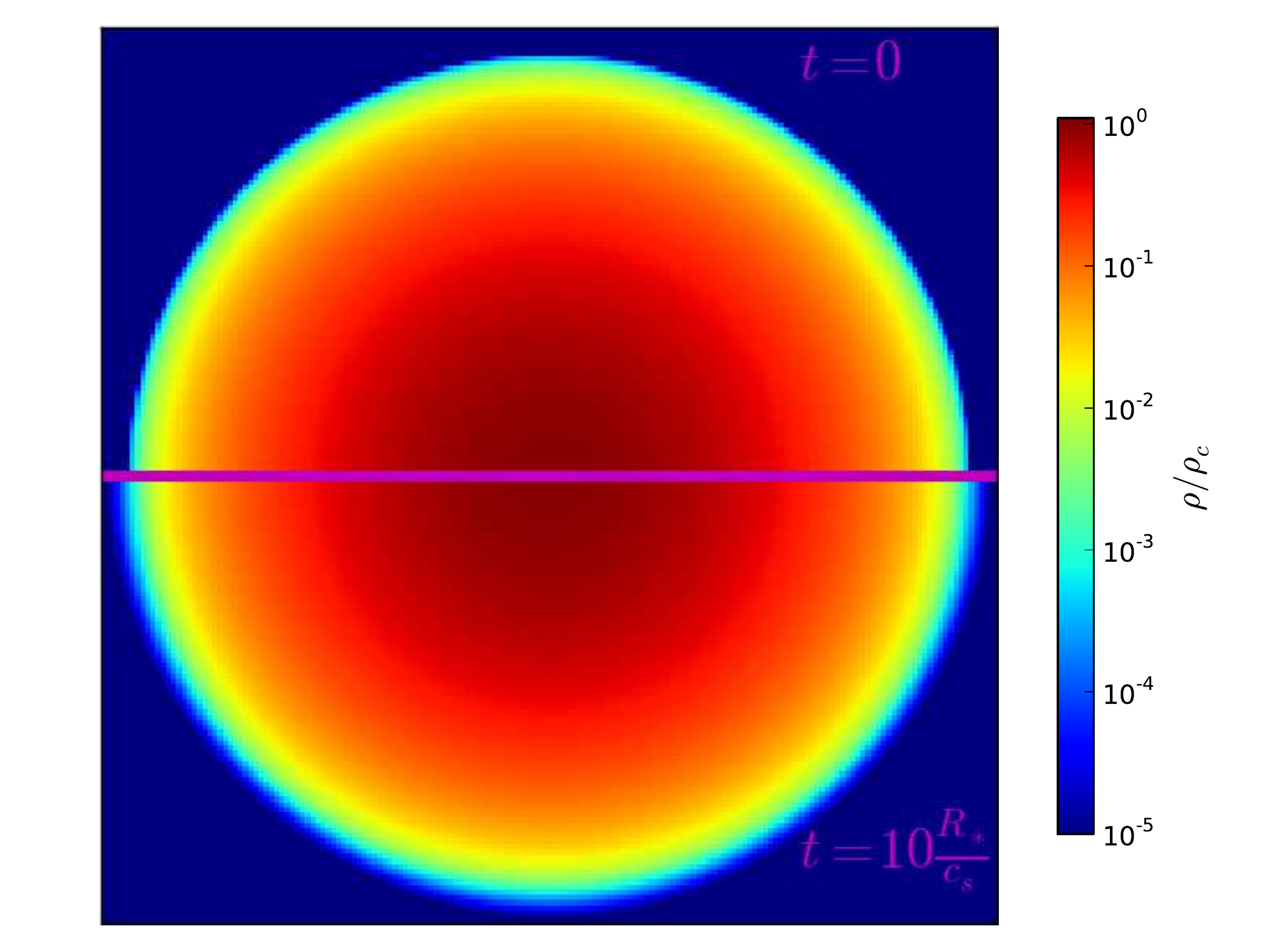}
\end{center}
\caption{
Snapshot of rest-mass density (in units of the initial central density of the star) for an isolated main-sequence star solution at $t=0$ (top) and after being evolved $10 R_*/c_s$ (bottom), where $c_s$
is the central sound speed, with resolution corresponding
to approximately 40 cells across the radius of the star on the finest level. This illustrates how well the simulation can maintain an equilibrium solution
at this resolution.
\label{star_eq}
}
\end{figure}

The simulations make use of adaptive mesh refinement based on truncation error.  Most of the simulations described here have up to eight levels of mesh
refinement with a 2:1 refinement ratio and a resolution of $\approx 0.01 M_{\rm BH}$ on the finest level (for the one $M_{\rm BH}/M_*=2\times10^6$ case, we add one additional level)
and a resolution of $ 1.2 M_{\rm BH}$ in the wave zone.
In Figure~\ref{conv_psi4_fig}, we also show the gravitational waveforms from 
simulations with higher ($1.5$ times) and lower ($\approx 0.8$ times) resolution to demonstrate the expected second-order convergence 
and to indicate the level of truncation
error.  We also demonstrate how well the simulation balances self-gravity against pressure for an isolated
main-sequence star solution in Figure~\ref{star_eq}, where we compare the initial solution to one evolved with the lower resolution for $10 R_*/c_s$, where $c_s$
is the sound speed at the center of the star. (For comparison, for the simulation with the main-sequence star and $M_{\rm BH}/M_*=10^6$, the radial free-fall time from the initial separation
of $d=200 M_{\rm BH}$ used here is $\approx 4 R_*/c_s$.)    
For the 3D simulations, the resolution is equivalent to the lower value of resolution, and since the same coordinates, etc., are used, the truncation
error is comparable.

\section{Cases}
\label{cases}
We consider two different types of stars.  To initially represent a main-sequence star, we choose the $\Gamma=5/3$ polytropic
Tolman--Oppenheimer--Volkoff solution with compaction $M_{*}/R_{*}=2\times10^{-6}$  (equivalent to $R_{*}\approx(M_*/M_{\odot}) R_{\odot}$ ).
To represent a white dwarf, we choose the $\Gamma=5/3$ polytropic solution with compaction $M_{*}/R_{*}=2.5\times10^{-4}$ 
(equivalent to $R_{*}\approx(M_*/M_{\odot}) 6\times10^3$ km ).  
As described in~\cite{1994ApJ...422..227C}, for such polytropic solutions one has the freedom to choose any overall
length scale (or equivalently, mass scale), though when considering the potential GW detectability in Section~\ref{detect}, we will consider different
specific choices of physical mass.

%
Because of the computational expense of the 3D cases, we restrict most of our simulations to head-on collisions.
We consider the head-on collisions of main-sequence stars with nonspinning BHs with mass ratios 
$M_{\rm BH}/M_{*}=1$, 0.5, 0.25, and 0.125 $\times 10^6$.  
The main-sequence star has a nominal Newtonian tidal disruption radius
of $r_T/M_{\rm BH}=50 \left( M_{*}/M_{\rm BH}\times{10^6} \right)^{2/3}$. 
(We note that in geometric units $10^6$ $M_{\odot}$ is equivalent to $\approx1.5\times10^6$ km or $\approx5$ s.)
In order to probe any effects of BH spin, we also consider 
the collisions of main-sequence stars with BHs with dimensionless spin $a=0.99$, where the spin and collision axes are aligned.
For the spinning cases, which have more high-frequency power, we consider a slightly higher range of mass ratios given 
by $M_{\rm BH}/M_{*}=2$, 1, 0.5, and 0.25 $\times 10^6$
We also study the head-on collisions of a white dwarf with a nonspinning BH with mass ratios 
$M_{\rm BH}/M_{*}=4$, 2, 1, and 0.5 $\times 10^3$.  The white dwarf has a nominal Newtonian tidal disruption radius of 
$r_T/M_{\rm BH}=40 \left( M_{*}/M_{\rm BH}\times{10^3} \right)^{2/3}$.
With these parameters, for the same value of the tidal disruption
radius in units of BH mass (i.e., the same value of $r_T/M_{\rm BH}$), the white dwarf's size compared to the BH (i.e., $R_*/M_{\rm BH}$) will be 
larger than for main-sequence stars.  Hence, we can anticipate that GW suppression due to decoherence will happen at somewhat smaller values
of $r_T/M_{\rm BH}$ for the white dwarf compared to the main-sequence star.  

Finally, we consider two of the more computationally expensive nonzero-angular-momentum collisions of a main-sequence star and a nonspinning BH with $M_{\rm BH}/M_{*}=10^6$.  These
cases have reduced orbital angular momentums of $\tilde{L}/M_{\rm BH}=2$ and $3.5$, which should be compared to the maximum
angular momentum for the star to fall into the BH of $\tilde{L}_{\rm cap}/M_{\rm BH}=4$.   
We do not consider the case with $\tilde{L}=\tilde{L}_{\rm cap}$ because tidal pancaking makes these simulations more computationally challenging, 
though we will argue in Section~\ref{geo_model} that decoherence is expected to strongly suppress the GW radiation in this case.

\section{Results}
\label{results}
\subsection{Main-sequence Stars}
\begin{figure}
\begin{center}
\includegraphics[width=3.6in,draft=false]{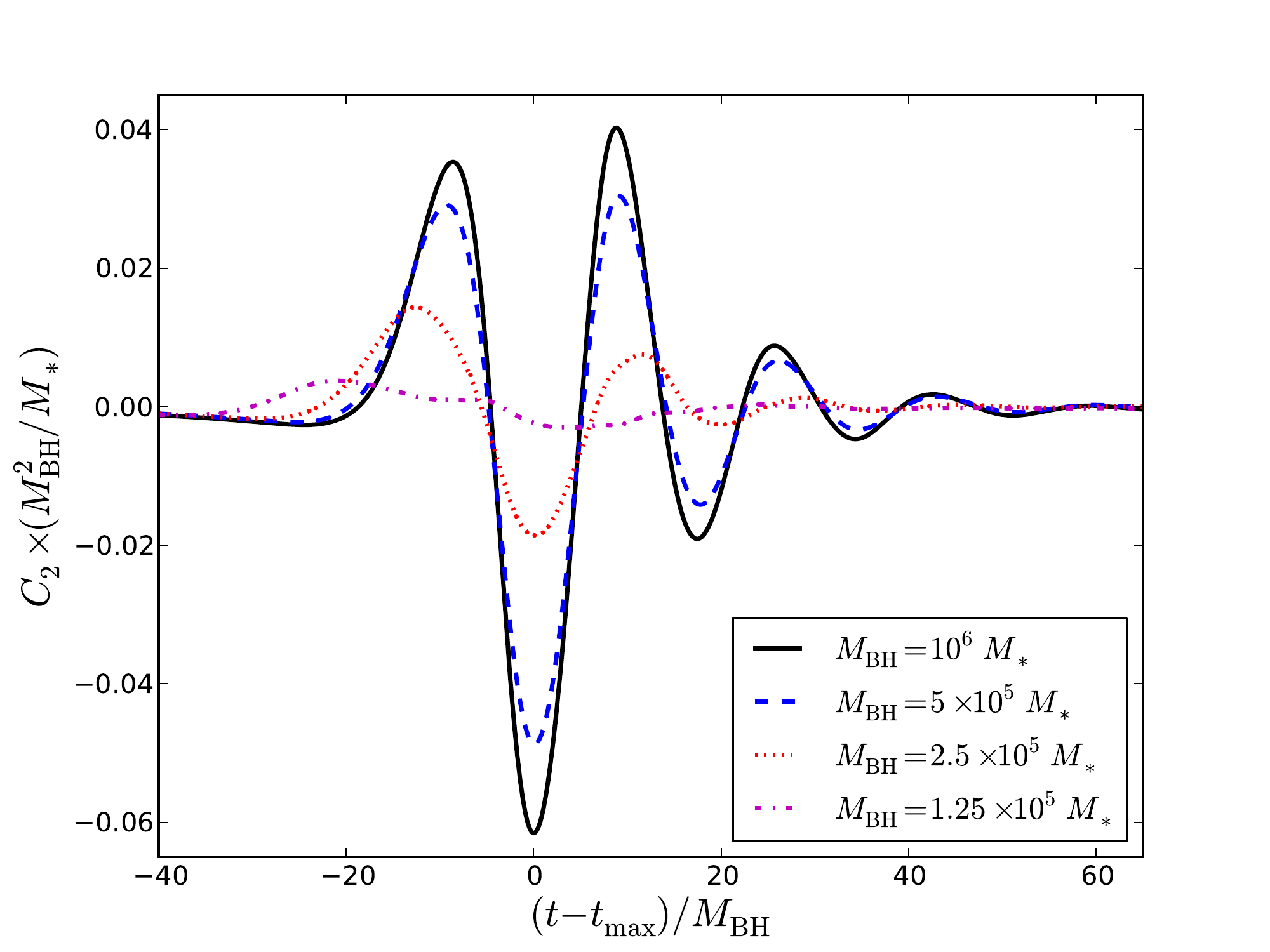}
\includegraphics[width=3.6in,draft=false]{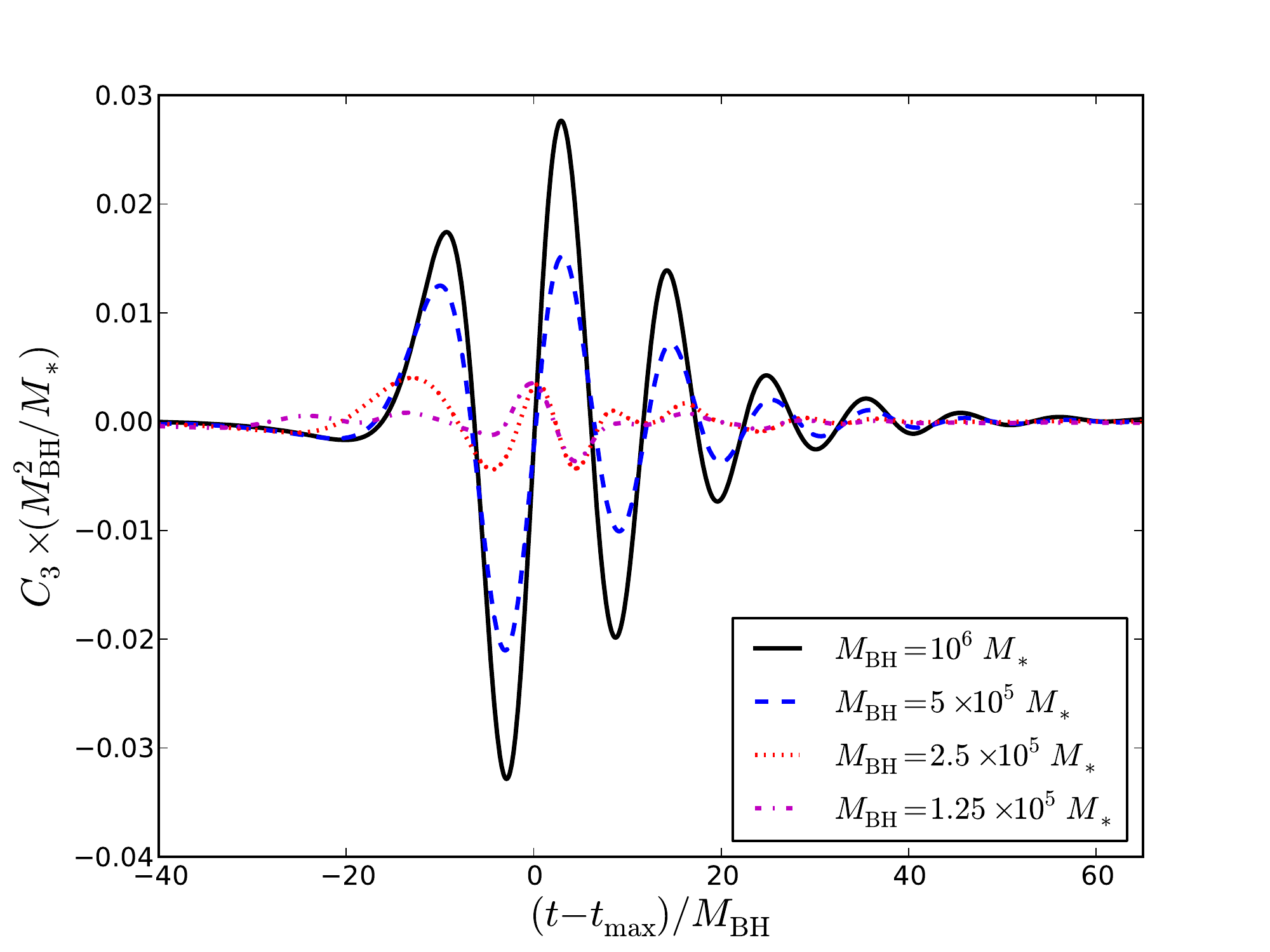}
\includegraphics[width=3.6in,draft=false]{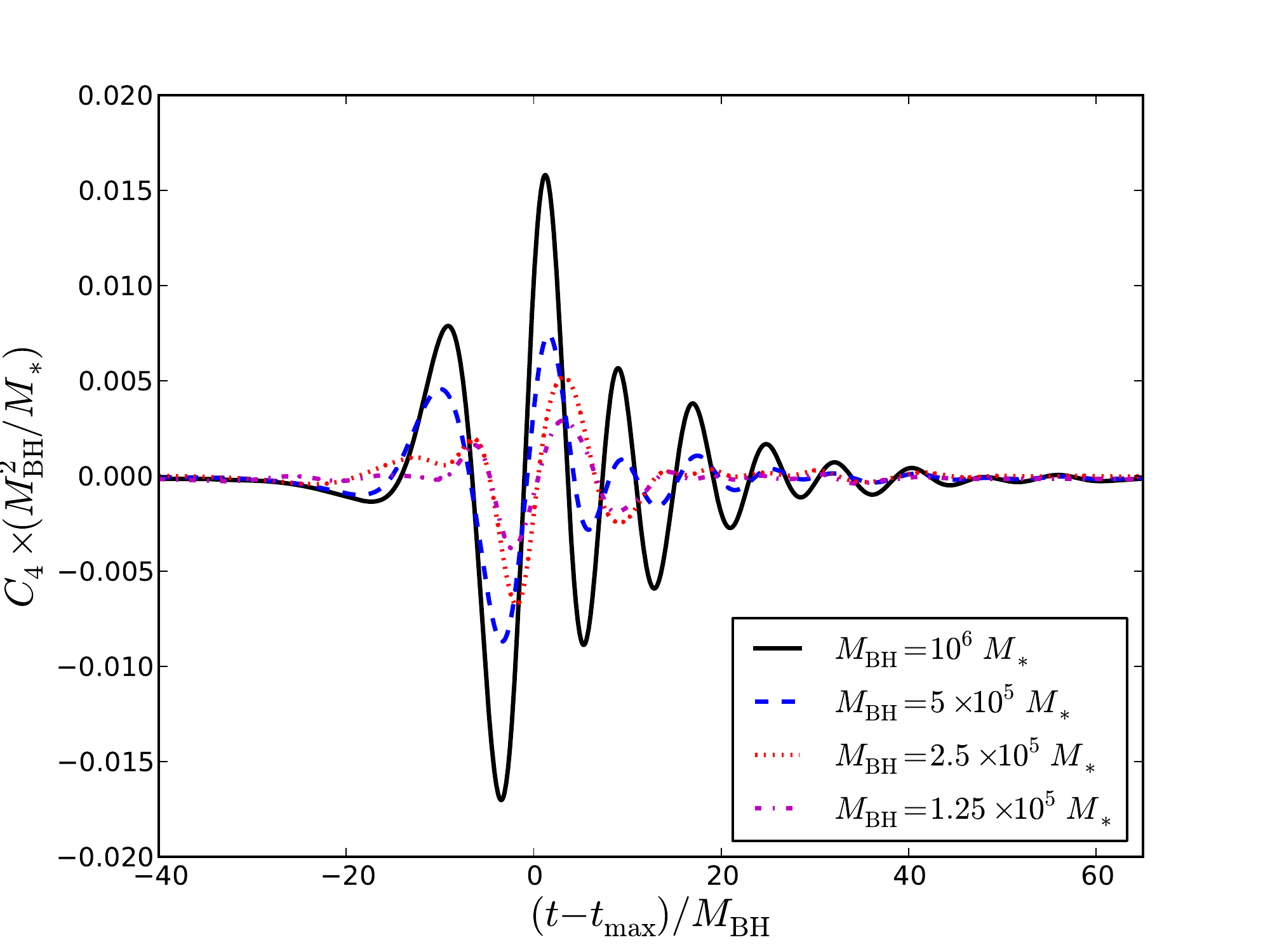}
\end{center}
\caption{
From top to bottom, the $\ell=2$, 3, and $4$ (with $m=0$) spin weight $-2$ spherical harmonics of $r\Psi_4$ for the head-on collision of a main-sequence star and a nonspinning BH for various mass ratios.
The amplitude has been scaled so that the curves would align in the point-particle limit.  Here $t_{\rm max}$ is the time when the $l=2$ amplitude is maximum for the largest mass ratio case.
\label{ns_psi4_fig}
}
\end{figure}

For star--BH collisions with low orbital angular momentum, the gravitational waveform will be dominated by the merger of the star with the BH, followed by
the quasi-normal mode (QNM) ringing of the perturbed BH post-merger.  In Figure~\ref{ns_psi4_fig}, we show the first three spherical harmonics of $\Psi_4$ from the head-on collisions of main-sequence stars with a BH, 
scaled with the mass ratio so that they would agree in the point-particle limit.
For $M_{\rm BH}/M_*\geq 10^6$, the waveform agrees to within a few percent (roughly comparable to the truncation error---see Figure~\ref{conv_psi4_fig}) with the 
point-particle prediction, which gives an energy in GWs of $E_{\rm GW}=0.0104 M_*^2/M_{\rm BH}$~\citep{Davis:1971gg}.  
Below this mass ratio,
the gravitational radiation begins to be suppressed. However, even at $M_{\rm BH}= 2.5\times10^5 M_*$,  the amplitude of the dominant $l=2$ mode is still $\sim 1/3$ of the point-particle value.  We emphasize that the waveforms are 
shown scaled by the mass ratio, and, for a fixed value of $M_{\rm BH}$, the amplitude of the 
$M_{\rm BH}= 2.5\times10^5 M_*$ simulation is in fact larger than the $M_{\rm BH}= 10^6 M_*$.  In the lower panels of Figure~\ref{ns_psi4_fig}, we can see that the higher harmonics
are slightly more suppressed at smaller mass ratios, which can be related to the fact that higher $\ell$ QNMs are at slightly higher frequencies (e.g., the least-damped QNM mode frequencies
are $\omega_{\rm QNM}M_{\rm BH}=0.37$, 0.60, and 0.81 for $l=2$, 3, and 4 ($m=0$), respectively~\citep{Berti:2005ys}
and are thus more easily suppressed when the differential timescale over which the different mass elements of the star merge with the BH is comparable 
to $M_{\rm BH}$.

\begin{figure}
\begin{center}
\includegraphics[width=3.6in,draft=false]{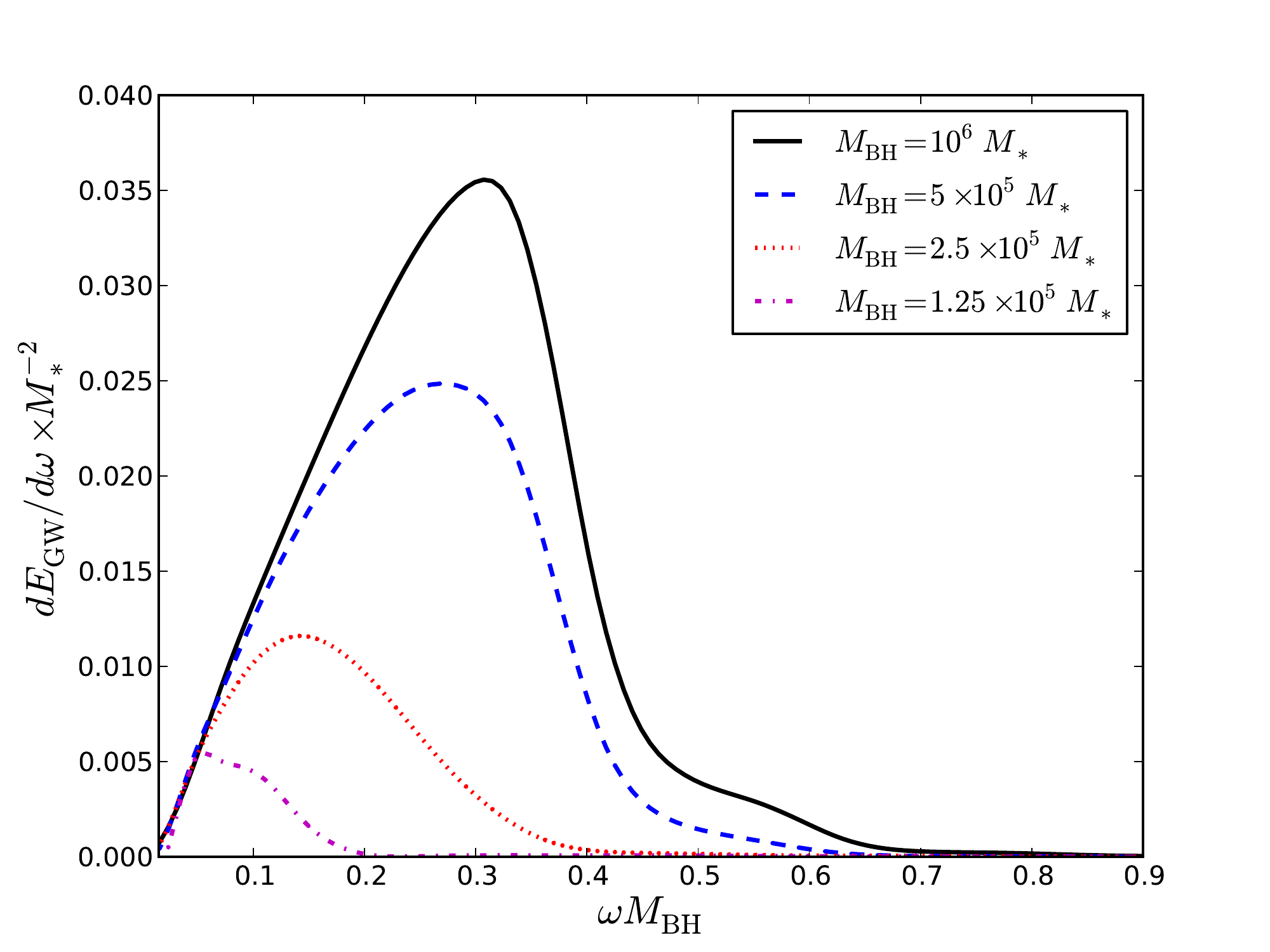}
\includegraphics[width=3.6in,draft=false]{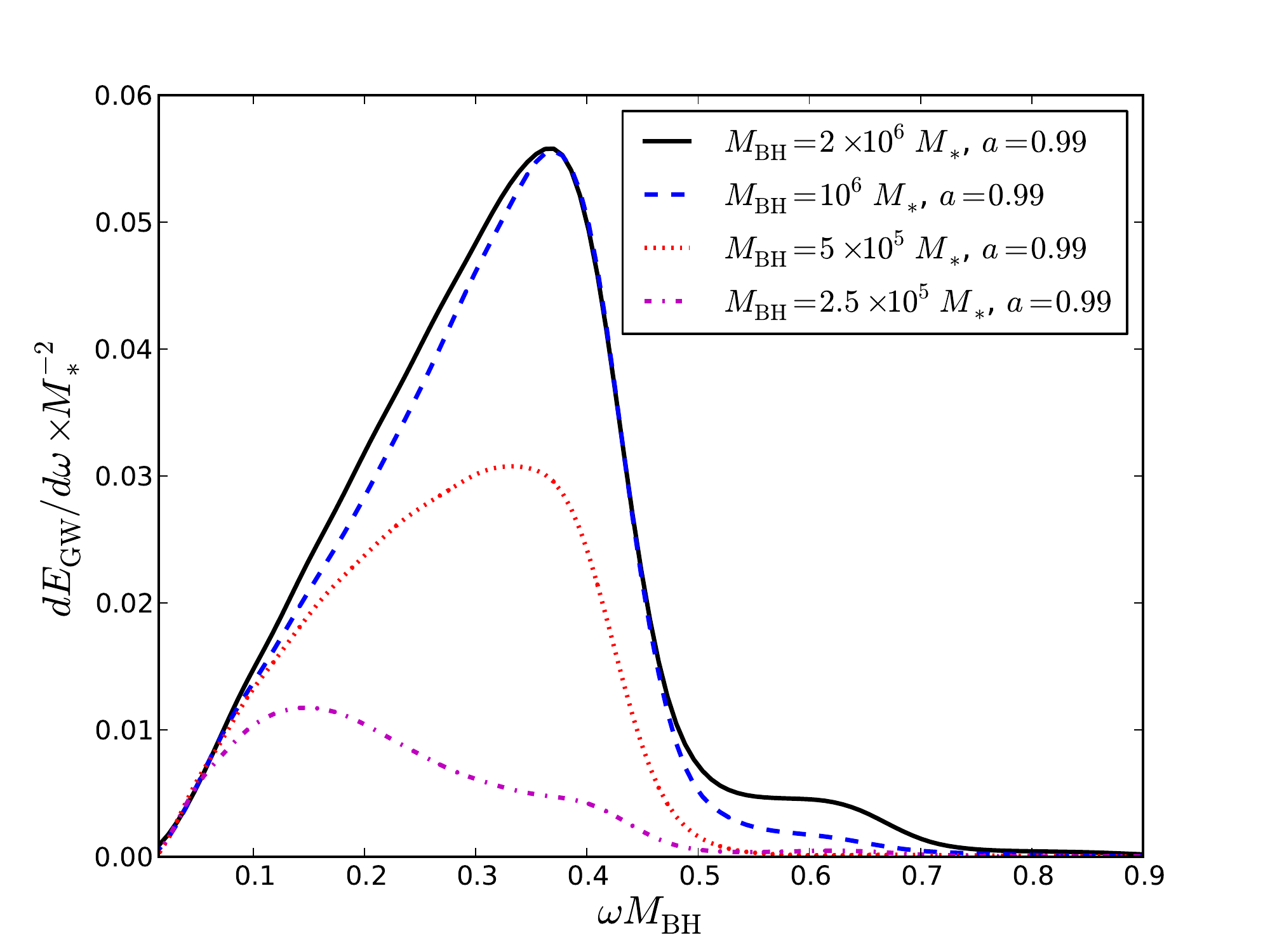}
\includegraphics[width=3.6in,draft=false]{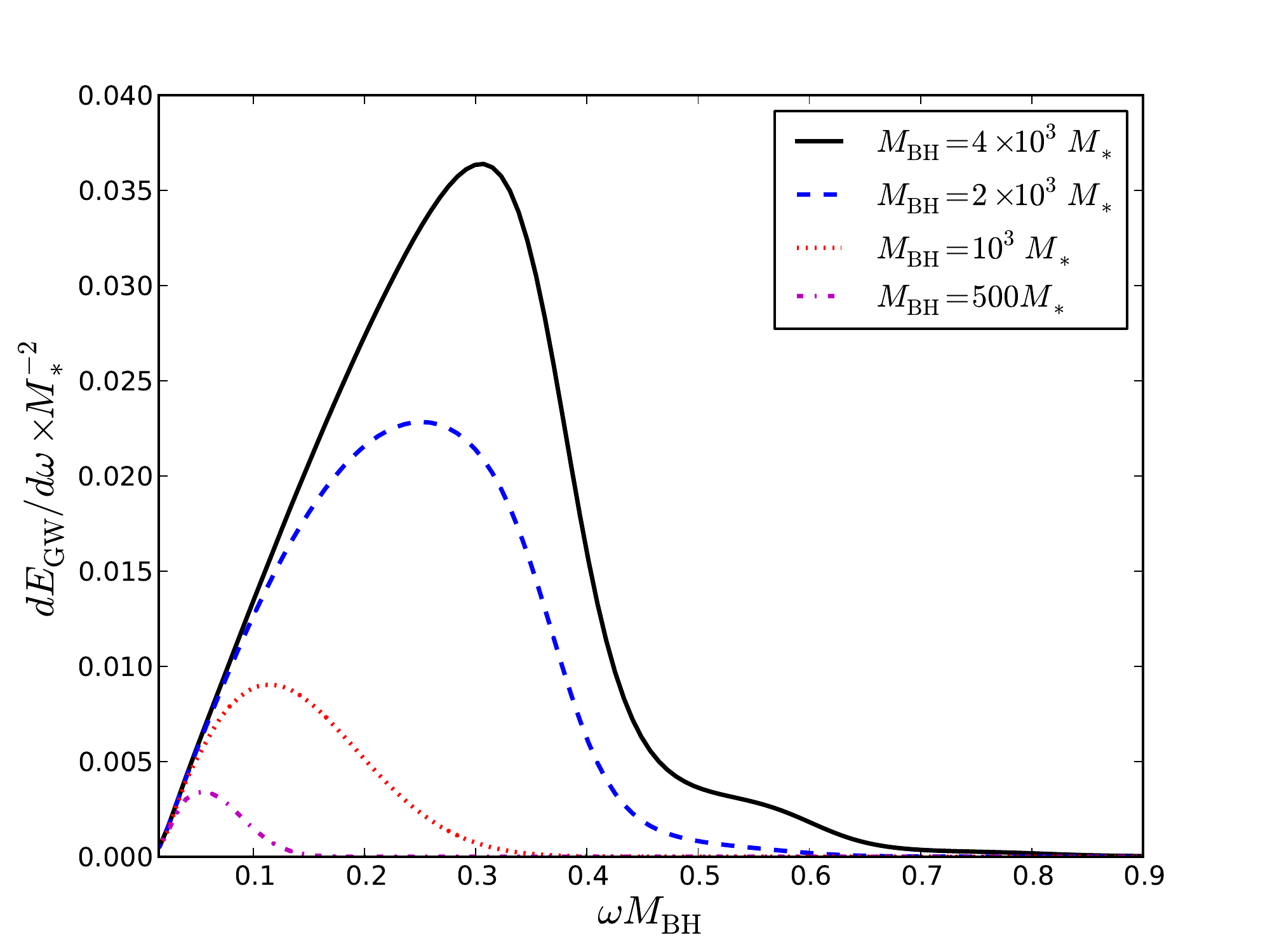}
\end{center}
\caption{
Gravitational wave power spectrum (scaled so that the curves would agree in the point-particle limit) from the head-on collision of a star and a BH. 
From top to bottom, we show results for a main-sequence star and a nonspinning BH, a main-sequence star and an $a=0.99$ BH, and a white dwarf and a nonspinning BH, all for various mass ratios.
\label{ns_egw_fig}
}
\end{figure}

In the top panel of Figure~\ref{ns_egw_fig}, we show the GW power spectrum for the same cases as in Figure~\ref{ns_psi4_fig}.  The low-frequency GW power 
comes from the star falling into the BH at large distances, while the spectrum peaks at around the BH's dominant QNM frequency,
 and falls off rapidly above.  From this figure, 
it is apparent that with decreasing mass ratio, and hence increasing tidal radius, more and more of the higher 
frequency power is suppressed.  This results in the peak power shifting to lower frequencies.

\begin{figure}
\begin{center}
\includegraphics[width=3.6in,draft=false]{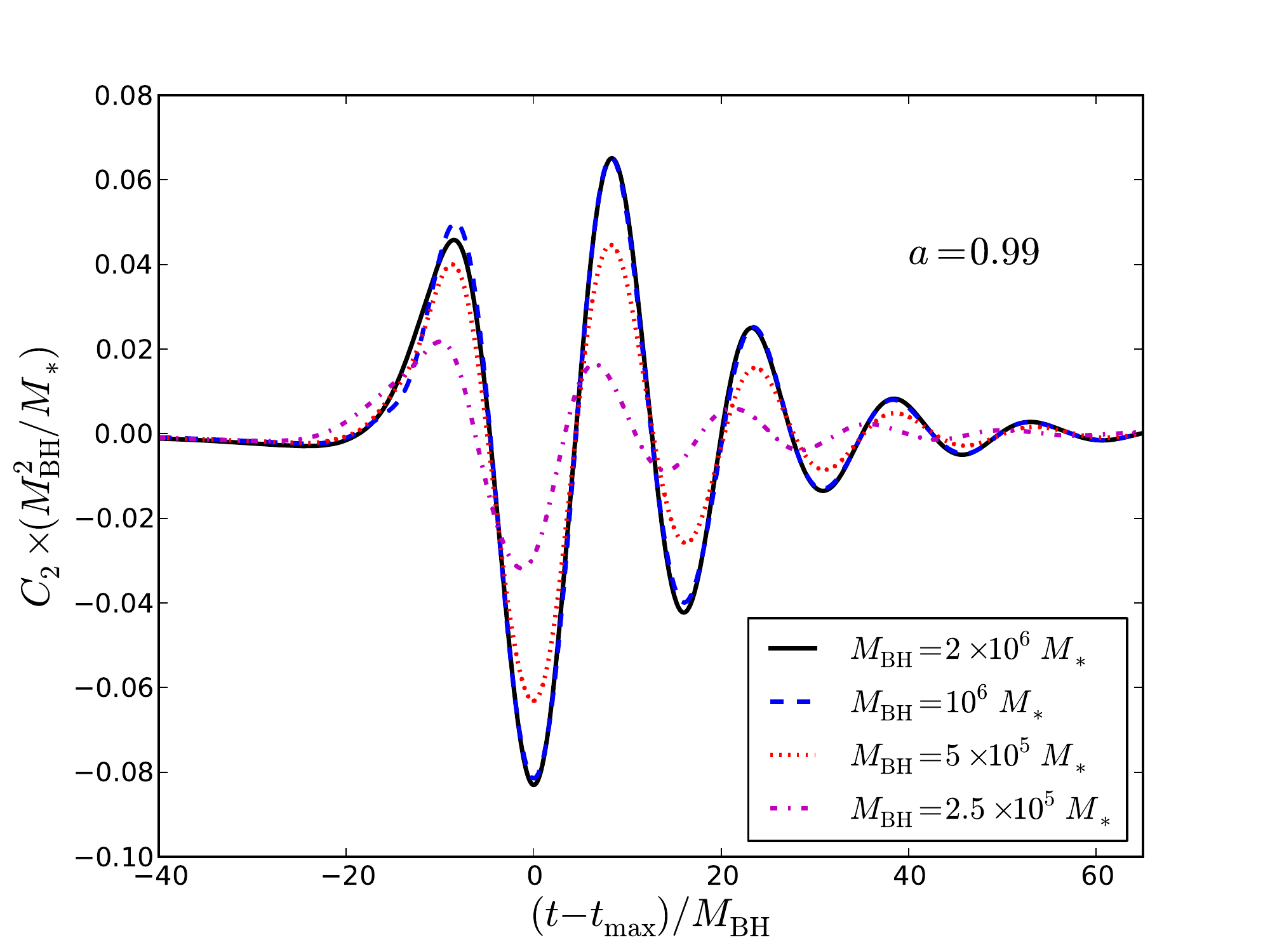}
\includegraphics[width=3.6in,draft=false]{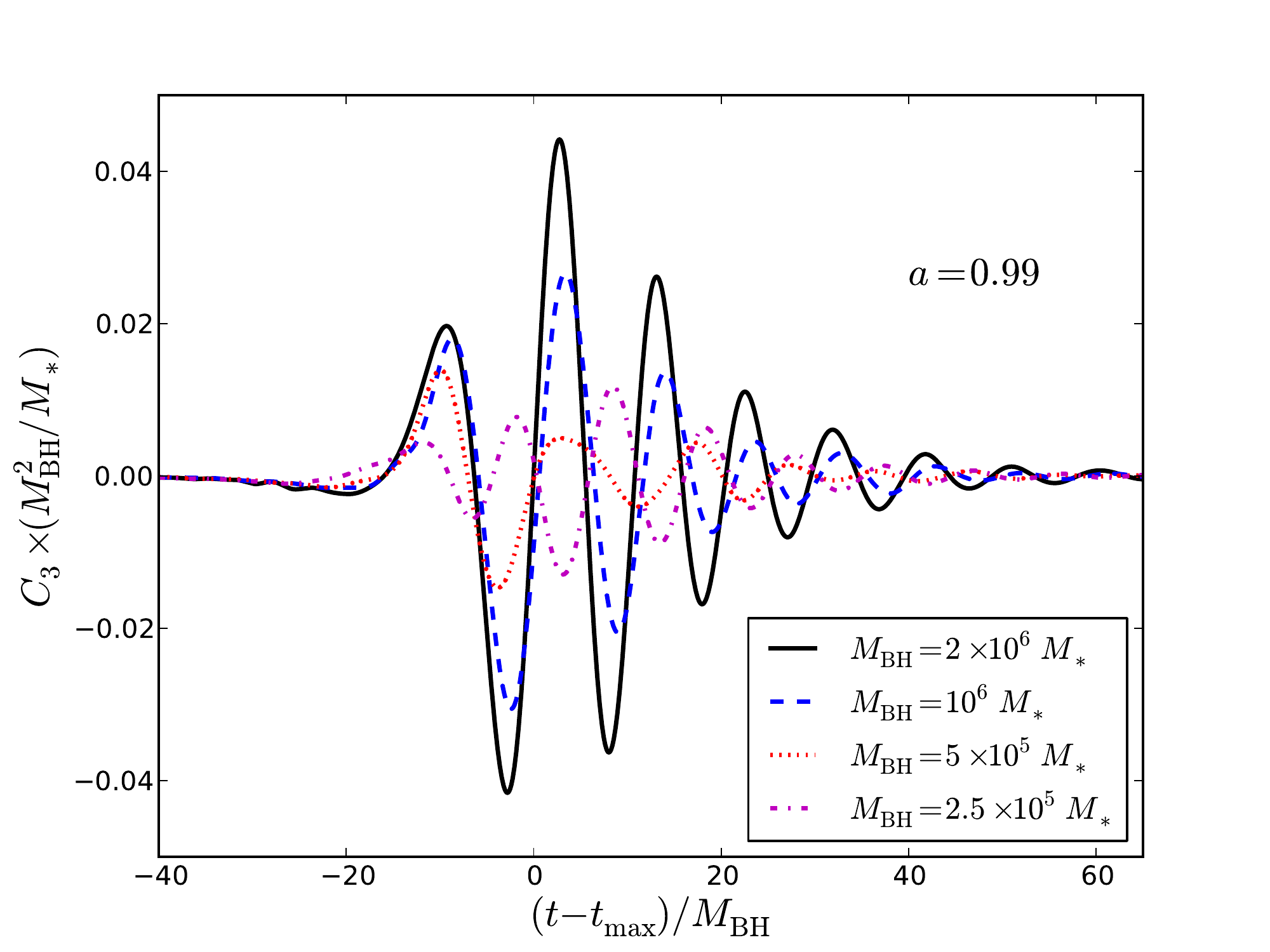}
\includegraphics[width=3.6in,draft=false]{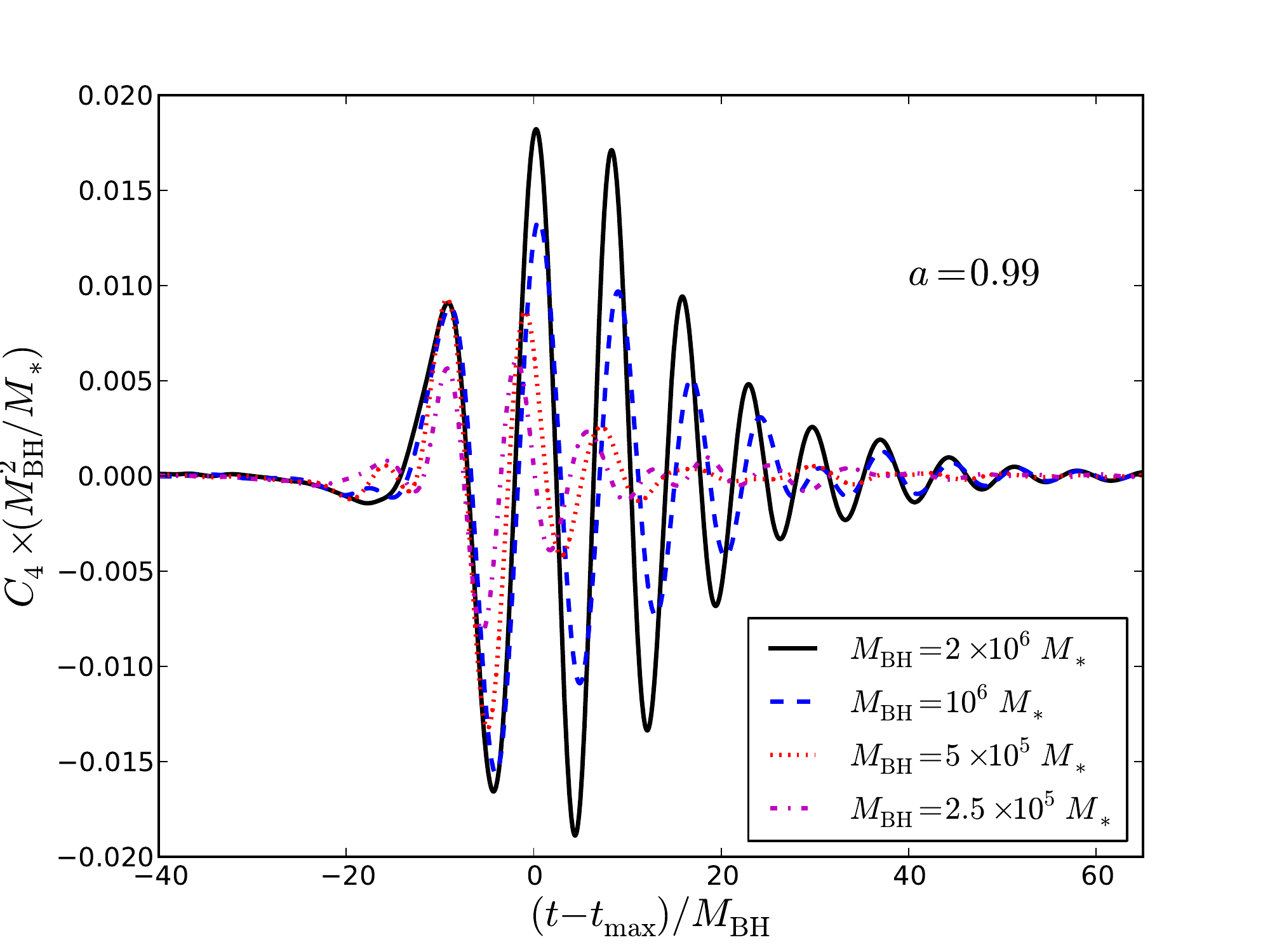}
\end{center}
\caption{
Same as Figure~\ref{ns_psi4_fig}, but for the head-on collision of a main-sequence star and a BH with dimensionless spin $a=0.99$.
\label{s_psi4_fig}
}
\end{figure}

We also perform the same simulations as above, but with a BH with dimensionless spin $a=0.99$.  We note that in order 
to preserve the axisymmetry of the simulation, we consider the case where the collision axis and the spin axis
are aligned.  In the point-particle limit, this setup results in $1.65$ times as much energy in GWs as the nonspinning
case~\citep{Nakamura1982185}, though only $0.4$ times as much energy as in the case where the spin and collision axes are 
perpendicular~\citep{Kojima1983335}.
In Figure~\ref{s_psi4_fig} and in the middle panel of Figure~\ref{ns_egw_fig}, we show,  respectively, the gravitational waveforms and power spectra
 for these simulations.  
These figures illustrate that the merger-ringdown of a highly spinning BH has additional power at higher frequencies.
The behavior with mass ratio is similar to the nonspinning case, though in this case the GW energy in the $M_{\rm BH}/M_*=10^6$ case is slightly more
suppressed at $\approx 85\%$ of the point-particle prediction. 
Comparing to the higher mass ratio $M_{\rm BH}/M_*=2\times10^6$ case, we can see that the former case is missing energy at higher frequencies, and that, in
particular, the higher $\ell$ modes (which resonant at higher frequencies) are noticeably suppressed. 

\begin{figure}
\begin{center}
\includegraphics[width=3.6in,draft=false]{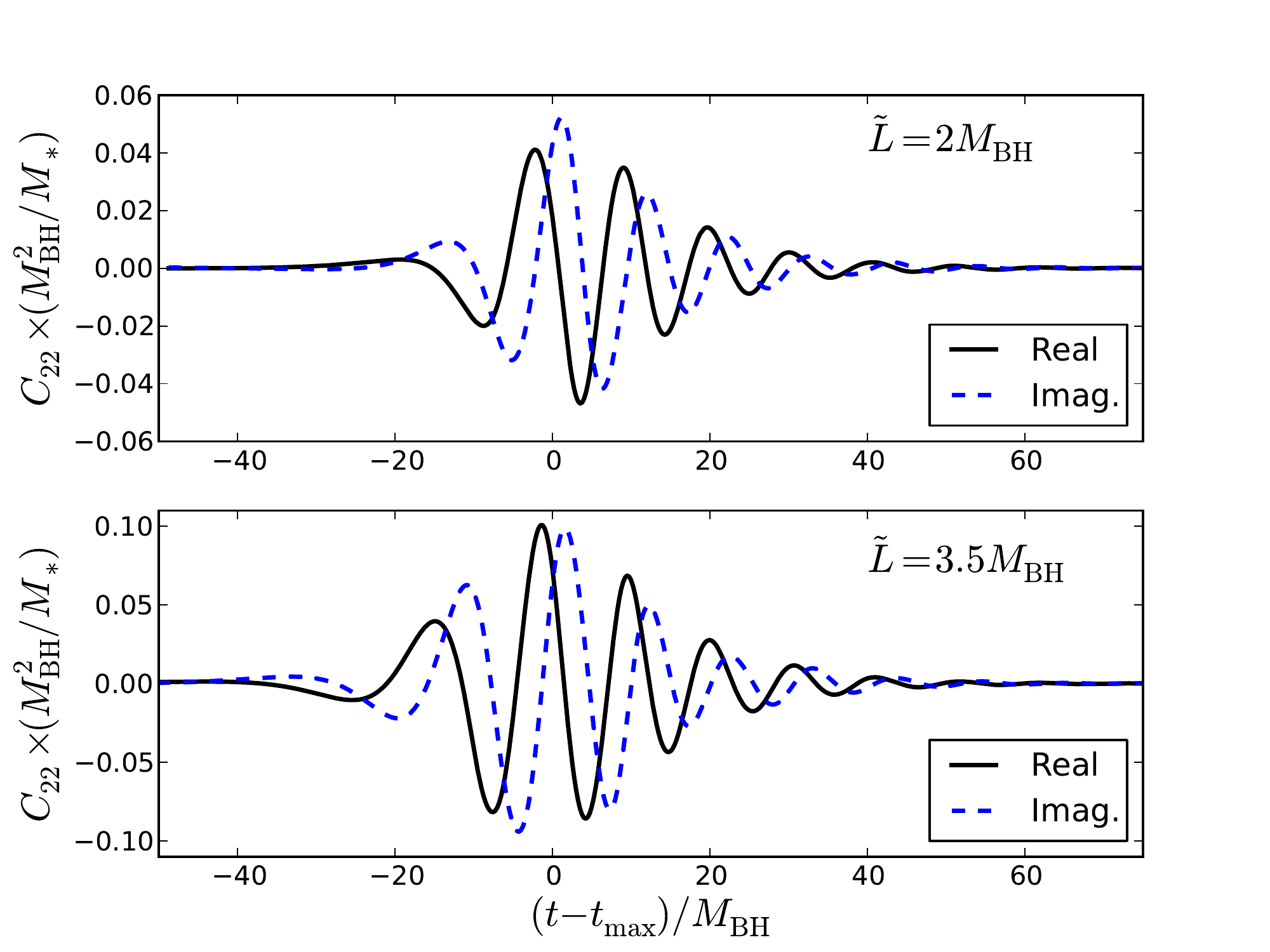}
\end{center}
\caption{
$\ell=m=2$ spin weight $-2$ spherical harmonic of $r\Psi_4$ for the collision of a main-sequence star and a BH with $M_{\rm BH}/M_*=10^6$
and $\tilde{L}/M_{\rm BH}=2$ (top) and 3.5 (bottom). 
\label{3d_psi4_fig}
}
\end{figure}

\begin{figure}
\begin{center}
\includegraphics[width=3.6in,draft=false]{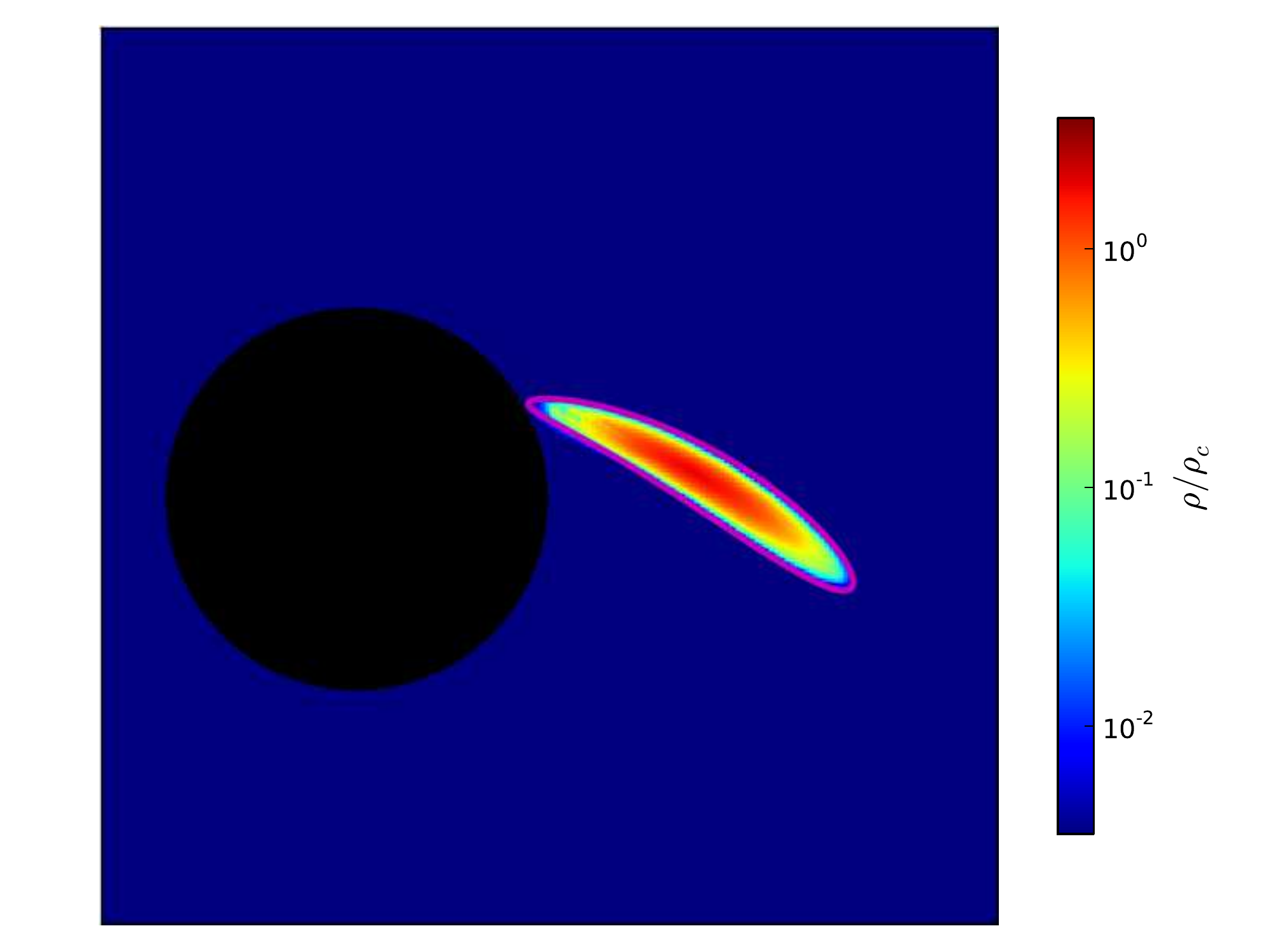}
\includegraphics[width=3.6in,draft=false]{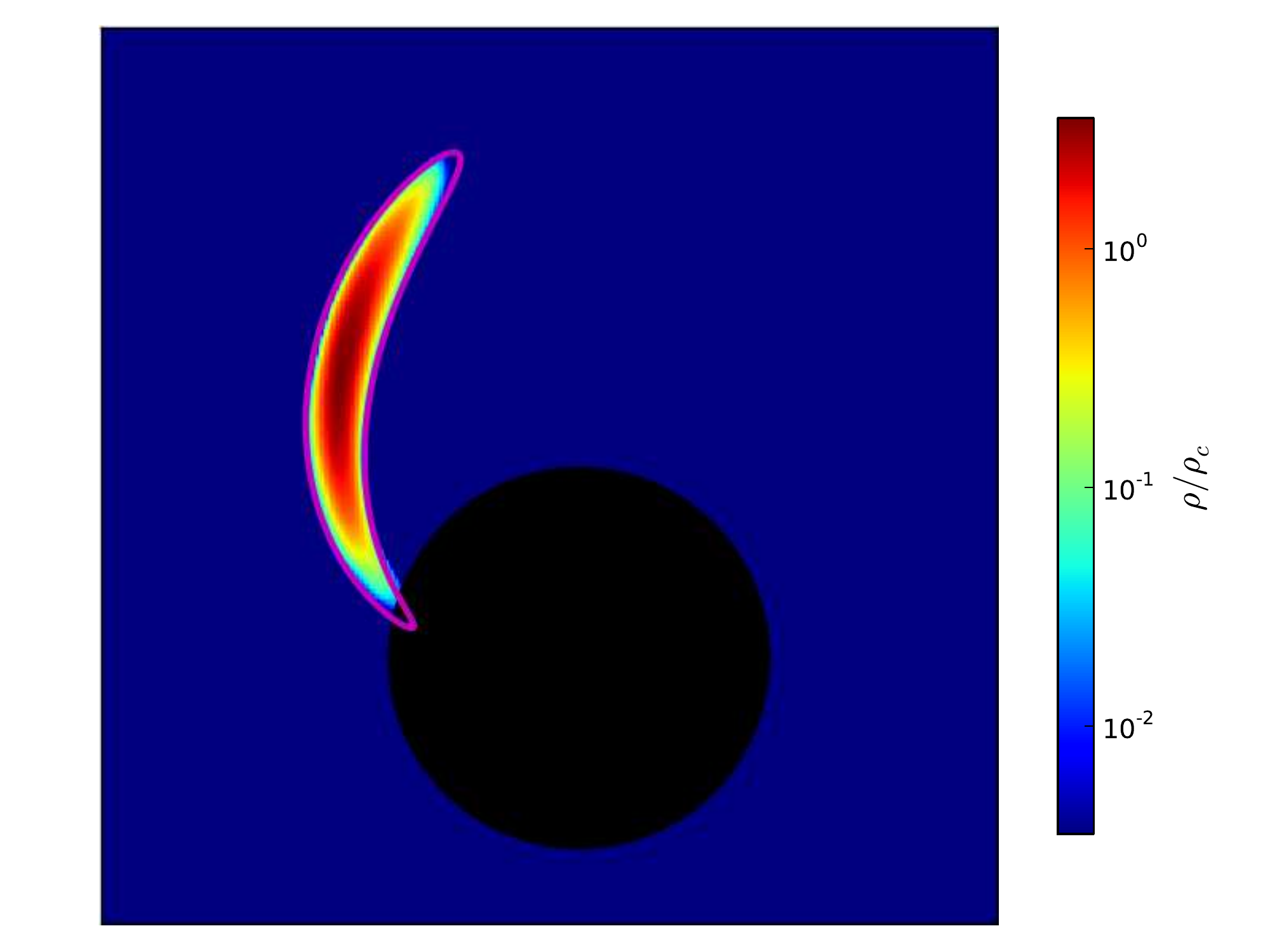}
\end{center}
\caption{
    Snapshots of rest-mass density (in units of the initial central density of the star) in the orbital plane around the time that the star's center of mass
    crosses the BH's light ring from simulations with main-sequence stars, $M_{\rm BH}/M_*=10^6$, and $\tilde{L}/M_{\rm BH}=2$ (top), and $3.5$ (bottom).
    Also shown is the position of the set of geodesics from the corresponding model described in Section~\ref{geo_model} (magenta outline) at the same time as the simulation snapshot.
    \label{rho_ms_fig}
}
\end{figure}

Coherent merger-ringdown GW signals are possible not only for head-on star--BH collisions, but also for those with nonnegligible
angular momentum.  In Figure~\ref{3d_psi4_fig}, we show the dominant $\ell=m=2$ component of the waveform for two cases with 
reduced angular momentum $\tilde{L}/M_{\rm BH}=2$ and $3.5$. These two cases have $E_{\rm GW} M_{\rm BH}/M_*^2\approx 0.039$ and $0.19$, respectively,
which correspond to $\sim90\%$ and $\sim75\%$ of the point-particle prediction. With increasing angular momentum, the waveform amplitude increases and 
begins to show evidence of ``whirling" behavior before plunging into the BH. Figure~\ref{rho_ms_fig} shows snapshots of the star's fluid density
from these cases shortly before collision.  As can be seen, with increasing angular momentum, the star also becomes more stretched out in the orbital plane, though
denser at its center due to being compressed in the perpendicular direction. However, for
$\tilde{L}/M_{\rm BH}=3.5$, the star's size is still comparable to the BH's.  
In Section~\ref{geo_model},
we estimate how much closer to $\tilde{L}_{\rm cap}$ the star may be able to coherently excite GWs.

\subsection{White Dwarfs}
\begin{figure}
\begin{center}
\includegraphics[width=3.6in,draft=false]{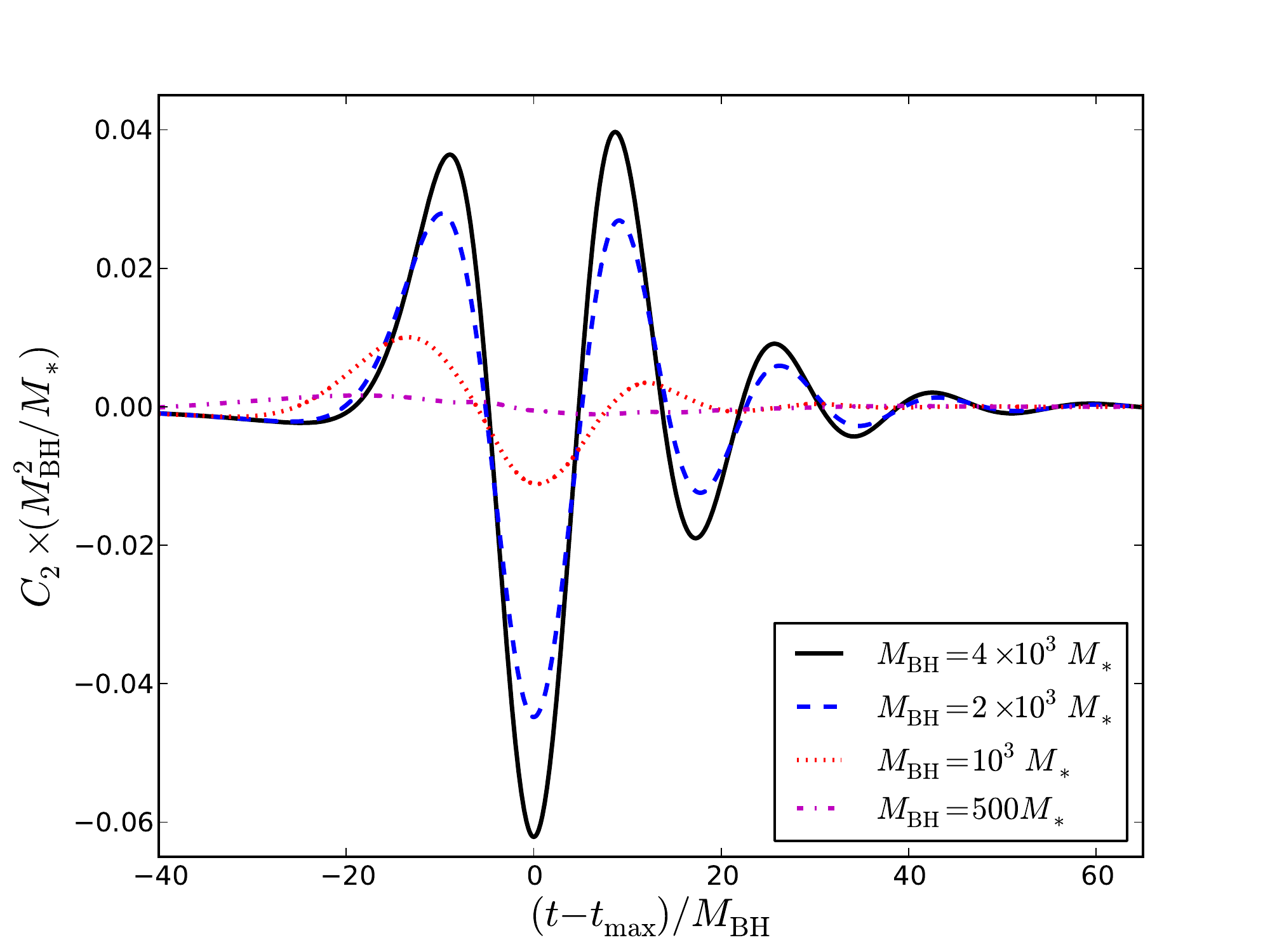}
\includegraphics[width=3.6in,draft=false]{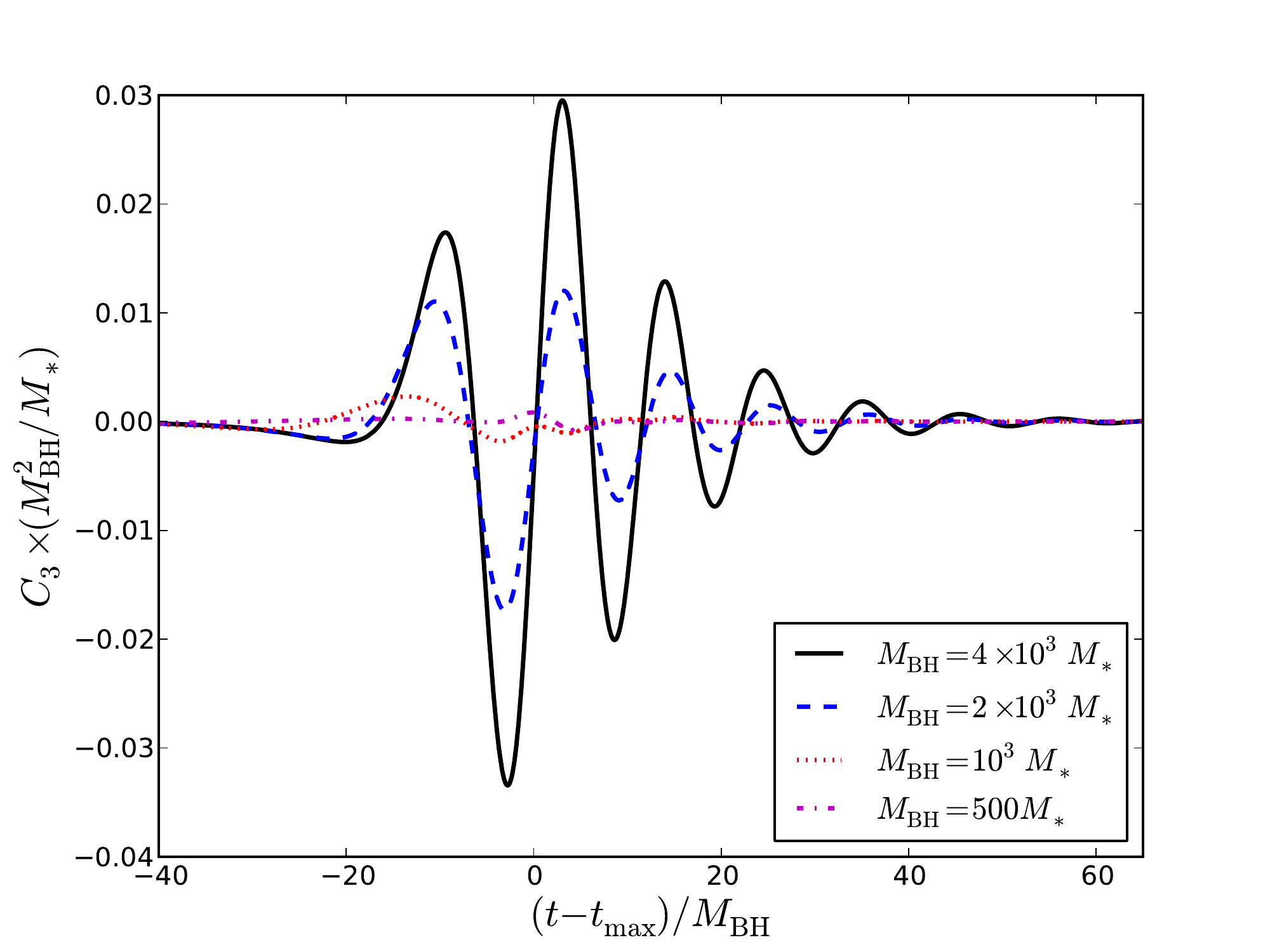}
\includegraphics[width=3.6in,draft=false]{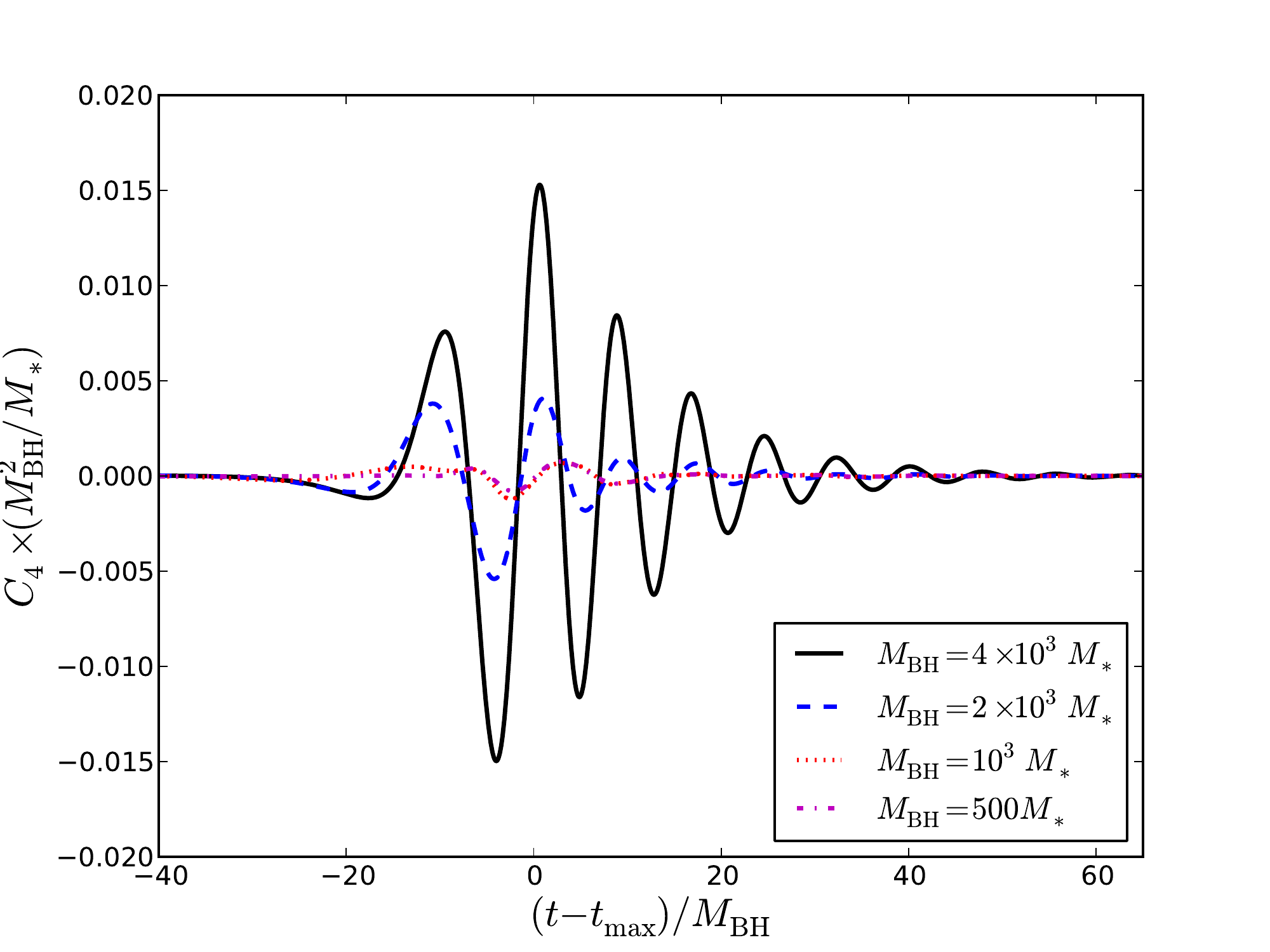}
\end{center}
\caption{
Same as Figure~\ref{ns_psi4_fig}, but for the head-on collision of a white dwarf and a nonspinning BH.
\label{wd_psi4_fig}
}
\end{figure}

\begin{figure}
\begin{center}
\includegraphics[width=3.6in,draft=false]{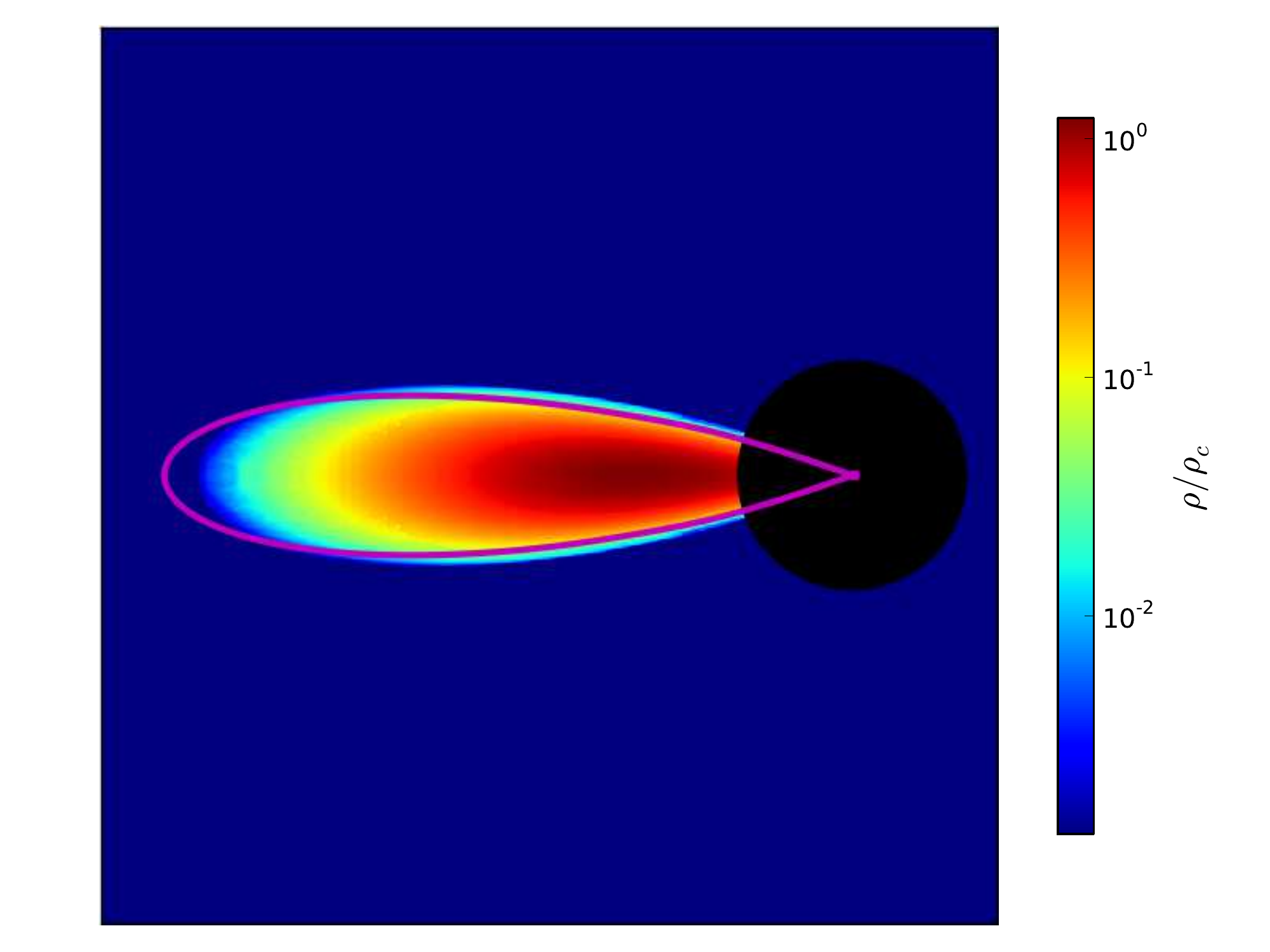}
\includegraphics[width=3.6in,draft=false]{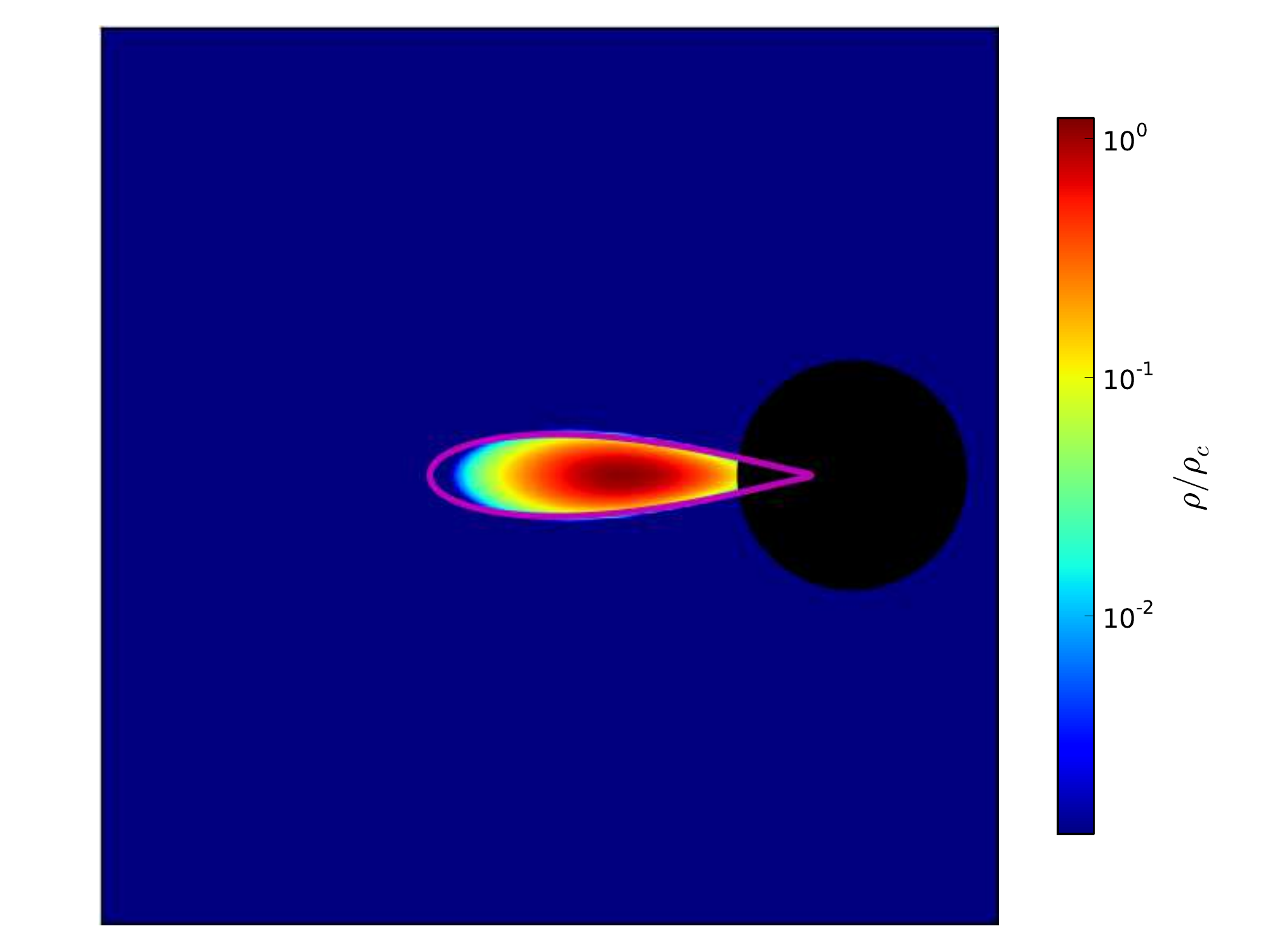}
\end{center}
\caption{
    Same as Figure~\ref{rho_ms_fig}, but for the head-on collision of a white dwarf with a BH with mass $M_{\rm BH}/M_*=2\times10^3$ (top) and $M_{\rm BH}/M_*=4\times10^3$ (bottom).
    \label{rho_wd_fig}
}
\end{figure}

For the collision of white dwarfs with massive BHs, we find qualitatively similar behavior to the main-sequence star 
cases around the transition from mass ratios that can be treated as close to point particles, to those where decoherence is important.
This can be seen in Figure~\ref{wd_psi4_fig} and the bottom panel of Figure~\ref{ns_egw_fig}.  When scaled
appropriately by the mass ratio, the waveform from the $M_{\rm BH}/M_{*}=4\times10^3$ case is almost
identical to the  $M_{\rm BH}/M_{*}=10^6$ main-sequence star case (and hence also the point-particle prediction).
For smaller mass ratios, this is no longer true. The $M_{\rm BH}/M_{*}=2\times10^3$ and $M_{\rm BH}/M_{*}=10^3$ cases
have, respectively, $\approx 65\%$ and  $\approx 15\%$ the energy in GWs (scaled by $M_{\rm BH}^2/M_*$) as the $M_{\rm BH}/M_{*}=4\times10^3$ case.
Compared to the main-sequence star simulations, the suppression of GW energy due to finite size effects not only sets in at smaller mass ratios 
(since the white dwarf is more compact) but also at somewhat smaller values of the tidal disruption radius. As noted above, this is because for the white-dwarf 
compaction used here, the white dwarf's unperturbed size is comparable to the radius of the BH at smaller values of $r_T$. 
This can be seen in Figure~\ref{rho_wd_fig} where we show snapshots of the white-dwarf density around the time of collision for the  $M_{\rm BH}/M_{*}=2\times10^3$ and $4\times10^3$
cases.

\section{Geodesic model of tidal disruption}
\label{geo_model}

In order to compare to the results from the full simulations, and in order to generalize
them to cases that were not run, we can construct a simple model based on geodesics of an isolated BH spacetime along the lines
considered in~\cite{1982ApJ...257..283H} and~\cite{2012PhRvD..85b4037K}.  Corresponding to a star with radius $R_*$
and tidal disruption radius $r_T$, we consider a set of equatorial geodesics parameterized by an angle $\theta$ 
with initial positions $(x,y)=(x_c+R_* \cos\theta,y_c+R_* \sin\theta)$ where $\sqrt{x_c^2+y_c^2}=r_T$ and all with 
the same initial velocity given by the trajectory of the star's center of mass at $r=r_T$.  Then we can compute the time $\Delta t_{\rm geo}$ 
it takes for these geodesics to cross the BH light ring compared to the center-of-mass geodesic.  
We ignore nonequatorial orbits since tidal forces will flatten the star in the direction perpendicular to the orbital plane.
This simplistic model of course ignores the dynamics of the disruption process and assumes that the combined effects of the star's self-gravity
and pressure support can be ignored for $r\leq r_T$, while, for $r>r_T$, tidal effects can be completely ignored.
It also does not capture the distribution of the star's mass, which, in general, will be centrally condensed.  
In addition, we note that the quantity $\Delta t_{\rm geo}$ is coordinate dependent (here we use harmonic coordinates as in the simulations).
Nevertheless, we find that this model captures the main features of the simulation results.

As an initial point of comparison, in Figures~\ref{rho_ms_fig} and~\ref{rho_wd_fig}, along with the density from the full simulation results at 
approximately the time the star's center of mass crosses the BH's light ring,
we also include the curve made up by the positions of the set of geodesics for the corresponding parameters and time.  We can see that in fact
this curve matches the shape of the star quite well in all cases.  This indicates that the model assumptions are fairly good approximations
for these cases.

\begin{figure}
\begin{center}
\includegraphics[width=3.6in,draft=false]{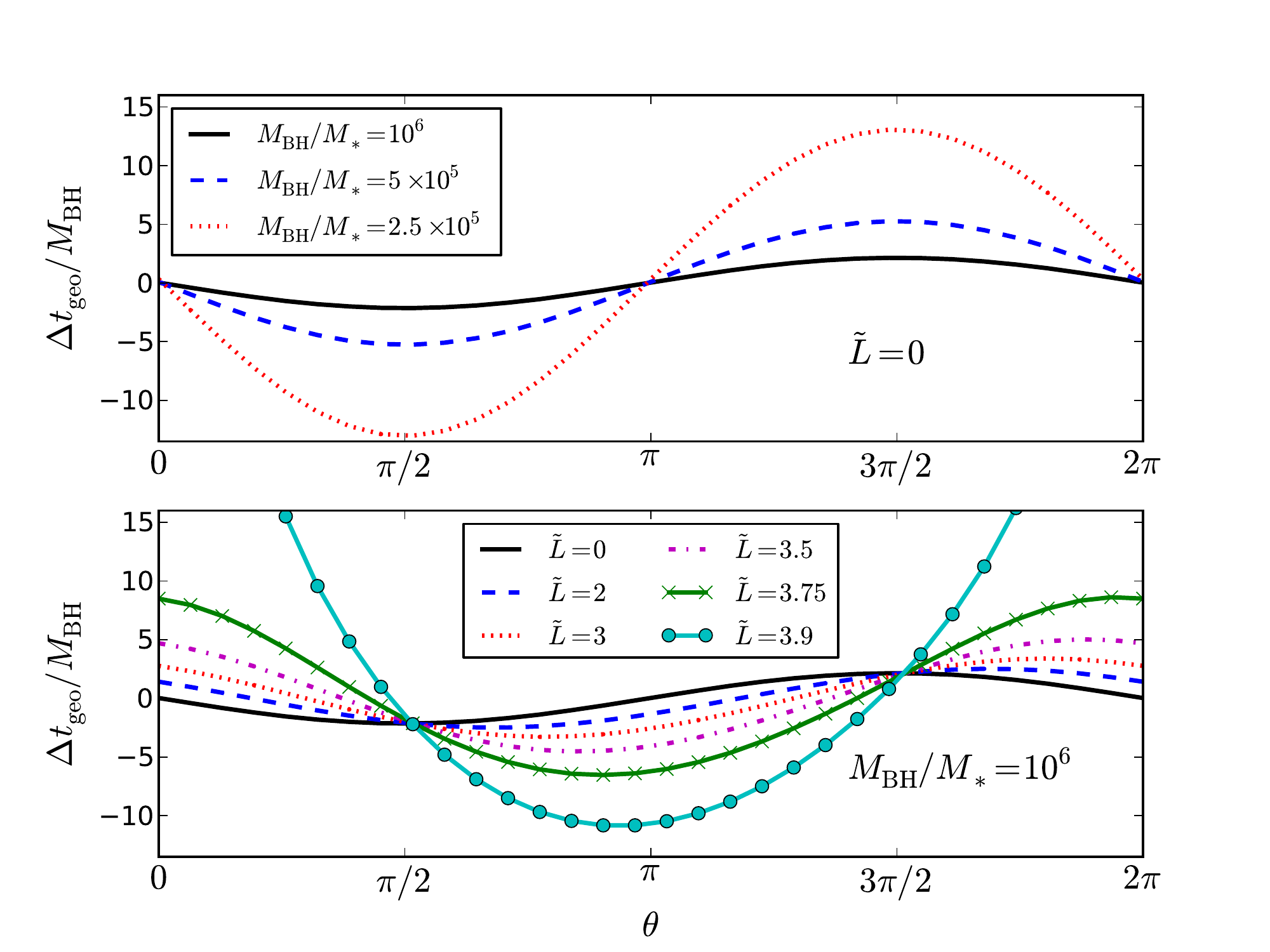}
\end{center}
\caption{
  Relative difference in time for geodesics to cross the BH light ring for model parameters corresponding to a main-sequence star.  
  The top panel corresponds to head-on collisions with different mass ratios, while the bottom panel corresponds to a $M_{\rm BH}/M_*=10^6$
  mass ratio with different amounts of angular momentum.
\label{geo_ms_fig}
}
\end{figure}

In Figure~\ref{geo_ms_fig}, we show 
the values of $\Delta t_{\rm geo}$ computed for different values of $R_*$ and $r_T$ corresponding to the model main-sequence star.
Since for large mass ratios the GW power spectrum peaks at $\omega \sim 0.3/M_{\rm BH}$ (Figure~\ref{ns_egw_fig}), we expect decoherence
of the GW signal when $|\Delta t_{\rm geo}|\gtrsim \pi/\omega \sim 10 M_{\rm BH}$.  The top panel of Figure~\ref{geo_ms_fig} shows that for head-on collisions
the $M_{\rm BH}/M_*=5\times10^5$ case lies below this threshold, while the $M_{\rm BH}/M_*=2.5\times10^5$ case exceeds it.  This is fully consistent with 
the results from the full simulations where the former case had $\sim 70\%$ of the point-particle prediction for GW energy, and the latter $\sim 25\%$.  
The bottom panel indicates that in the $\tilde{L}/M_{\rm BH}=2$ and $3.5$ cases with $M_{\rm BH}/M_*=10^6$,
gravitational radiation should not be significantly suppressed, also consistent with what was found in the simulations.  Furthermore, the geodesic model
suggests that this should hold for $\tilde{L}/M_{\rm BH}\lesssim 3.75$.  As $\tilde{L}\to \tilde{L}_{\rm cap}$, part of the star will collide with the BH, while part
of it will go back out on a long (possibly unbound) orbit, as illustrated by the  $\tilde{L}/M_{\rm BH}=3.9$ case in Figure~\ref{geo_ms_fig}.
Hence decoherence effects will be strong.

\begin{figure}
\begin{center}
\includegraphics[width=3.6in,draft=false]{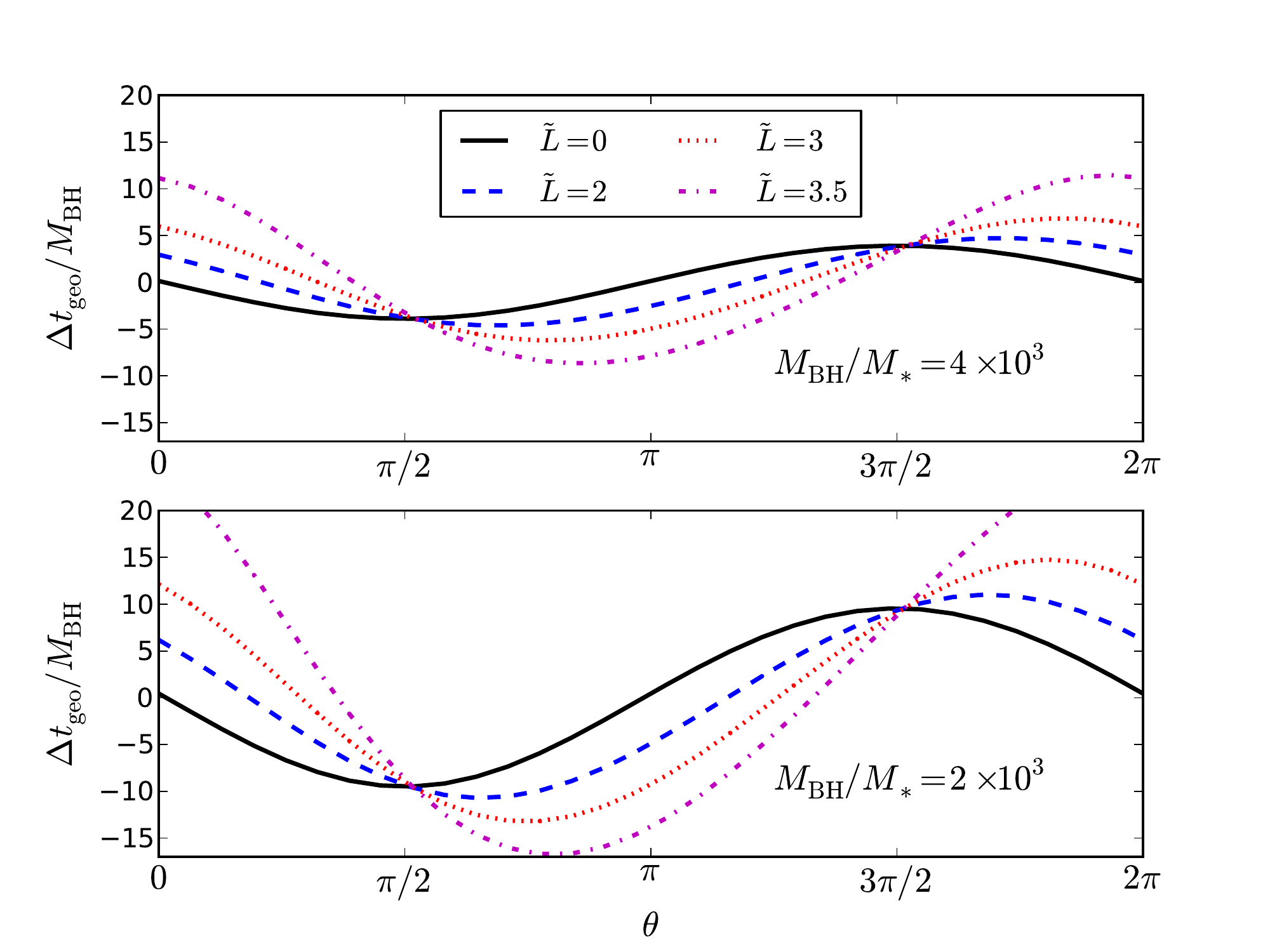}
\end{center}
\caption{
  Relative difference in time for geodesics to cross the BH light ring for parameters corresponding to a white dwarf.
  The top panel and bottom panel have parameters corresponding to mass ratios of $M_{\rm BH}/M_*=4\times10^3$ and 
  $M_{\rm BH}/M_*=2\times10^3$, respectively. 
\label{geo_wd_fig}
}
\end{figure}

We also apply the geodesic model to white dwarf parameters to obtain the results shown in Figure~\ref{geo_wd_fig}.
For $M_{\rm BH}/M_*=4\times10^3$ this indicates that decoherence should not set in for $\tilde{L}/M_{\rm BH}\lesssim 3.5$.
For the $M_{\rm BH}/M_*=2\times10^3$ case, the model indicates that a head-on collision should be right at the decoherence threshold
(we recall that the simulation for this case had $\approx 65\%$ of the point-particle prediction for GW energy), but that the magnitude of $\Delta t_{\rm geo}$
is only slightly increased for $\tilde{L}/M_{\rm BH}\leq 2$.

\section{Prospects for detection}
\label{detect}
\begin{figure}
\begin{center}
\includegraphics[width=3.6in,draft=false]{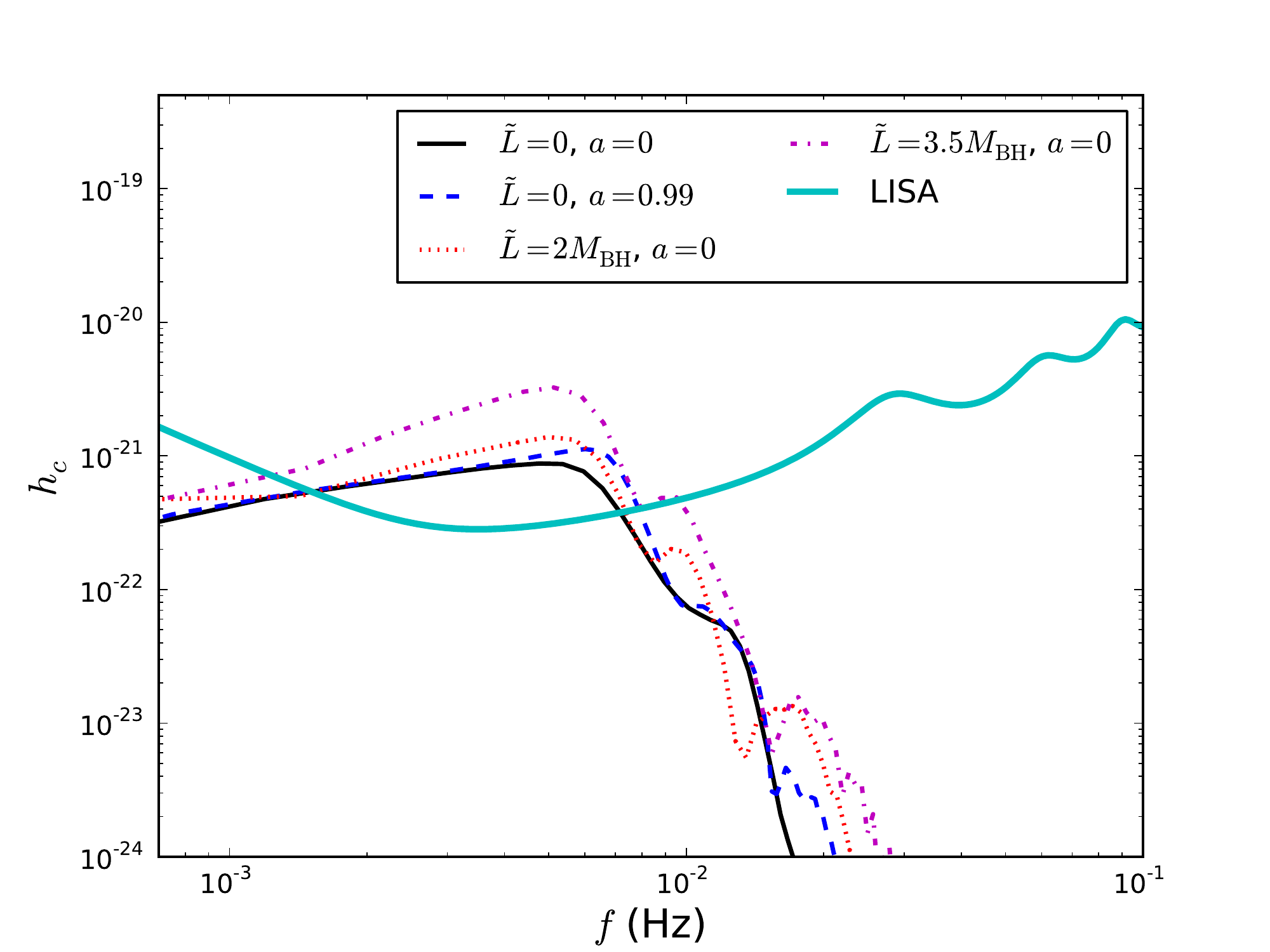}
\end{center}
\caption{
    Characteristic GW strain for main-sequence star--BH collisions with $M_*=2$ $M_{\odot}$ and $M_{\rm BH}=2\times10^6$ $M_{\odot}$ 
    assumed to be observed with optimal orientation at a distance of $d=15$ Mpc.  For comparison, we also show a proposed \emph{LISA} noise curve.  
    \label{detect_3d_fig}
}
\end{figure}

We can use the gravitational waveforms obtained from these simulations to estimate the distance at which future
GW detectors might be able observe such events.  To do so, we need to choose an overall physical
mass for the system to scale the results of the simulation. (We note that the choice $M_*/R_*=2\times10^{-6}$
for a main-sequence star is appropriate for $M_*\sim M_{\odot}$ and more massive stars will be slightly more
compact, while less massive stars will be less compact, but we ignore that weak dependence here.)
In Figure~\ref{detect_3d_fig}, we show the characteristic GW strain $h_c$ from the four simulations with 
$M_{\rm BH}/M_*=10^6$, assuming $M_*=2$ $M_{\odot}$ and an optimally oriented and located source at a distance of 15 Mpc.
The characteristic strain is defined as $h_c=|\tilde{h}|f$ where $\tilde{h}$ is the Fourier transform of the strain and
$f$ is frequency.  For comparison, we also show a proposed \emph{LISA} noise curve\footnote{We use the \emph{LISA} noise curve from~\cite{lisa_noise}  
that includes both pointing and shot noise, but remove the all-sky and polarization averaging factor as in~\cite{Berti:2005ys}. 
We also note that the proposed \emph{eLISA} design 
is less sensitive at lower frequencies than the original \emph{LISA} design~\citep{2013GWN.....6....4A}.} where for a noise power spectral density 
$S_n(f)$, we define the characteristic strain as $h_n=\sqrt{S_n f}$.  With these definitions, the signal-to-noise
of a match filtered signal is given by S/N$=2\left[\int h_c/h_n d\ln(f)\right]^{1/2}$, i.e., is given by integrating the ratio of characteristic strain signal- 
to-noise over a logarithmic frequency interval.  In Figure~\ref{detect_3d_fig}, we see that BHs with spin have an additional signal at high frequencies 
compared to nonspinning BHs, and that signals from collisions with more angular momentum are significantly stronger. For example, for $\tilde{L}/M_{\rm BH}=3.5$,  S/N$\approx 18$, 
while the $\tilde{L}/M_{\rm BH}=2$ and head-on case have, respectively, $\sim 0.5$ and $0.4$ the S/N of the head-on case for these parameters.
Non-head-on collisions with spinning BHs or collisions with somewhat higher $\tilde{L}$ than considered here
(the results of Section~\ref{geo_model} suggest $\tilde{L}/M_{\rm BH}\leq3.75$ will not be strongly suppressed) could potentially produce stronger signals.
The signal-to-noise is not extremely sensitive to physical mass in this range.  Scaling the results to $M_*=4$ $M_{\odot}$ instead of $M_*=2$ $M_{\odot}$ 
increases the S/N by $\sim70\%$ (this doubles the overall amplitude, but 
shifts the signal to lower frequencies where the sensitivity is somewhat lower), while scaling to $M_*=1$ $M_{\odot}$ roughly halves the S/N. 

\begin{figure}
\begin{center}
\includegraphics[width=3.6in,draft=false]{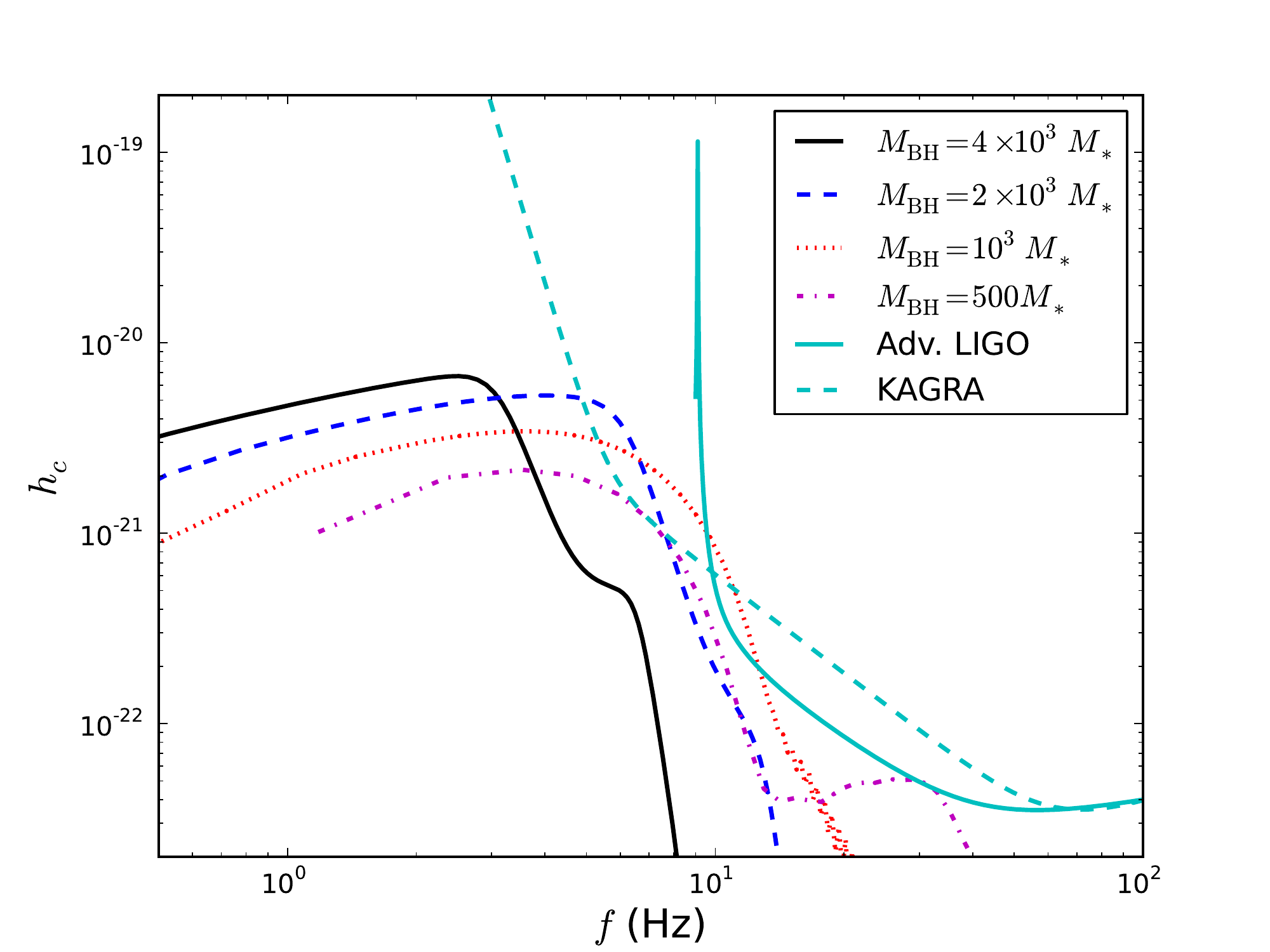}
\end{center}
\caption{
Characteristic GW strain for head-on white dwarf--BH collisions with $M_*=M_{\odot}$ and various values of $M_{\rm BH}$, 
assumed to be observed with optimal orientation at a distance of $d=1$ Mpc.  For comparison, we also show the Advanced LIGO and KAGRA noise curve.  
\label{detect_wd_fig}
}
\end{figure}

The frequency of the GWs produced by the white dwarf--BH systems considered here would fall at higher frequencies,
outside the range of a \emph{LISA}-like instrument.  The ringdown of BHs with masses $\lesssim 10^3$ $M_{\odot}$ will occur
in the Advanced LIGO sensitivity range; however, though those with higher masses will fall in the intermediate regime where
neither instrument is sensitive. Gravitational wave detectors utilizing atom interferometers could potentially fill this frequency gap~\citep{2008PhRvD..78l2002D}, 
as could other proposed space-based laser interferometers such as DECIGO~\citep{2006CQGra..23S.125K} or BBO~\citep{bbo}.
In Figure~\ref{detect_wd_fig}, we show the characteristic GW strain from the white-dwarf simulations, 
assuming $M_{*}=M_{\odot}$ and observation with optimal orientation at a distance of $d=1$ Mpc.
For comparison, we also show the proposed broadband Advanced LIGO noise curve~\citep{ligo_noise} and the broadband KAGRA noise curve~\citep{kagra_noise}.
We can see that for this particular choice of physical mass, the strongest signal occurs for $M_{\rm BH}/M_{*}=$ 1 to 2 $\times 10^3$ (S/N$\approx3$ for the KAGRA noise curve) since
for higher BH mass the frequency of the signal is too low, and for lower masses, tidal effects strongly suppress the GW signal.
The curves shown are all for head-on collisions, though, of course, generically the collision will occur with angular momentum.
As mentioned in Section~\ref{geo_model}, the geodesic model for the $M_{\rm BH}/M_{*}=2\times 10^3$ mass ratio suggests 
a collision with $\tilde{L}/M_{\rm BH}\sim 2$ should be similarly coherent as the head-on case, thus angular momentum
could boost the detectability by a factor of a few.  A collision with a spinning BH would also be somewhat easier to detect because of the 
extra power at high frequencies.

\section{Conclusion}
\label{conclusion}
In this paper, we explored the GWs produced by collisions of strongly tidally affected 
stars with massive BHs.  Taking into account the dynamics of the tidally perturbed star collision
process and the distribution of the star's mass, we found that nonnegligible GWs are produced well into the regime
where the star is colliding with the BH over a timescale comparable to the period of the GWs.  
We also found that collisions with angular momentum can substantially boost the GW strength, as can BH spin.
The main features of our results were well-captured by a simple model based on geodesics.

For main-sequence stars and BHs with $M_{\rm BH}\geq 10^6M_{*}$, we found that for most of the range of merging values of
angular momentum, the waveform is still similar to that of a point particle. 
Tidal disruption event rates per galaxy are estimated to be roughly in the range of $\sim10^{-5}$--$10^{-3}$ yr$^{-1}$~\citep{2004ApJ...600..149W,2008ApJ...676..944G,2011ApJ...741...73V,2012PhRvD..85b4037K}
and to vary inversely with BH mass.  The results found here suggest that collisions occurring in the
Virgo cluster, which contains $\sim10^3$ galaxies, could be visible by a \emph{LISA}-like GW detector for a range of parameters.
While event rates involving intermediate-mass BHs are not well-known, our results also suggest that there is a small window of parameter
space for a white dwarf collision where the BH is not so massive that the ringdown signal falls outside the frequency range of a ground-based detector, but still not
small enough in mass that decoherence completely suppresses the gravitational radiation.
While these sources will most likely not have very high rates for near-future GW detectors, these results do illustrate
that such events can be direct probes of the BH's characteristics if they occur sufficiently close, or
if they are targeted by a subsequent generation of detectors.    

Though neutron stars have significantly higher compactions than the cases considered here,
these results might also be used to inform an understanding of the GW signal produced when
a neutron star merges with a stellar-mass BH and how tidal effects may suppress the merger-ringdown
signal.  
For future study, it would also be interesting to explore high-angular-momentum collisions with spinning 
BHs since the point-particle prediction suggests that spin could enhance the strength of 
the GW signal significantly~\citep{Kojima198337}.
In upcoming work, we will use these same numerical methods to explore noncollisional tidal disruption events~(W.~E.~East \& F.~Pretorius 2014, in preparation).

\acknowledgments
I thank Frans Pretorius and Nick Stone for valuable comments on this work.
Simulations were run on the {\bf Bullet} cluster at SLAC and the {\bf Orbital} cluster 
at Princeton University.

\bibliographystyle{hapj.bst}
\bibliography{bhstar}

\end{document}